\documentclass[conference]{IEEEtran}

\ifCLASSINFOpdf
\else
\fi
\usepackage{dirtytalk}
\usepackage{multirow}
\usepackage{graphicx}
\usepackage{balance}
\usepackage{hyperref}

\usepackage{etoolbox}
\patchcmd{\quote}{\rightmargin}{\leftmargin 1.5em \rightmargin}{}{}

\usepackage{tcolorbox}
\AtBeginEnvironment{tcolorbox}{\small}

\begin{document}
%
\title{Privacy Starts with UI: Privacy Patterns and Designer Perspectives in UI/UX Practice}

\author{\IEEEauthorblockN{Anxhela Maloku, Alexandra Klymenko, Stephen Meisenbacher, and Florian Matthes}
	\IEEEauthorblockA{Technical University of Munich\\ School of Computation, Information and Technology\\Garching, Germany\\
		\{anxhela.maloku, alexandra.klymenko, stephen.meisenbacher, matthes\}@tum.de}}


\IEEEoverridecommandlockouts
\makeatletter\def\@IEEEpubidpullup{6.5\baselineskip}\makeatother
\IEEEpubid{\parbox{\columnwidth}{
		Symposium on Usable Security and Privacy (USEC) 2026 \\
		27 February 2026, San Diego, CA, USA \\
		ISBN 978-1-970672-07-7 \\
		https://dx.doi.org/10.14722/usec.2026.23079 \\
		www.ndss-symposium.org, https://www.usablesecurity.net/USEC/
}
\hspace{\columnsep}\makebox[\columnwidth]{}}

\maketitle

\begin{abstract}
In the study of Human-Computer Interaction, privacy is often seen as a core issue, and it has been explored directly in connection with User Interface (UI) and User Experience (UX) design. We systematically investigate the key considerations and factors for privacy in UI/UX, drawing upon the extant literature and 15 semi-structured interviews with experts working in the field. These insights lead to the synthesis of 14 primary design considerations for privacy in UI/UX, as well as 14 key factors under four main axes affecting privacy work therein. From these findings, we produce our main research artifact, a UI/UX Privacy Pattern Catalog, which we validate in a series of two interactive workshops and one online survey with UI/UX practitioners. Our work not only systematizes a field growing in both attention and importance, but it also provides an actionable and expert-validated artifact to guide UI/UX designers in realizing privacy-preserving UI/UX design.
\end{abstract}


%
\IEEEpeerreviewmaketitle

\section{Introduction}
In recent years, the unmistakable coupling of humans and computer interfaces has continued to pervade, most notably with advancements in the capability and accessibility of modern LLM systems. While evidence exists that humans are growing more aware and wary of how their personal data is collected, used, and sometimes
exploited in such systems \cite{mcclain2023americans}, other studies suggest that this awareness is still often stunted or even manipulated via the use of deceptive patterns in User Interface and User Experience (UI/UX) design \cite{bongard2021definitely,gray2021end}. It has therefore become clear that the design of the UI represents a critical point in the study of Human-Computer Interaction (HCI) and the deployment of user-facing systems -- especially where concerns of privacy are at play.

The specific consideration of \textit{privacy} in UI/UX has entered the research sphere, often in the realm of either \textit{privacy compliance} or \textit{dark patterns} \cite{gray2018uxside,mathur2021}. The view of privacy protection under the lens of compliance may frequently refer to the implementation of principles such as \textit{data minimization} in the design of software systems, including in UI/UX, and this is often guided by frameworks such as Privacy by Design (PbD) \cite{cavoukian2012privacy, wong2019}. In direct contrast to following proactive privacy-preserving guidelines such as PbD, \textit{dark patterns} refer to deceptive or manipulative design techniques in UI/UX that prioritize data collection or utility over user autonomy, which can directly compromise a user's privacy.

In this way, we find increasing amounts of evidence that privacy is crucial to UI/UX, yet there is relatively little systematization in UI/UX beyond the view of compliance, which often is generalist and remains purposely abstract, and dark patterns, which include but do not solely encompass privacy matters. Importantly, we find a lack of works focusing directly on the considerations that must be made by UI/UX designers and the specific contextual factors that affect these; moreover, no works do this with a foundation in perspectives from UI/UX experts and practitioners facing the technical, organizational, and ethical realities that privacy in UI/UX design brings. 

We address this gap by conducting a systematic study on the important considerations and factors for privacy in UI/UX design, which is grounded in three primary research questions:

\begin{enumerate}
    \item[RQ1.] What are the key privacy considerations for UI/UX design, and how can they be effectively addressed?
    \item[RQ2.] What are the key factors influencing UI/UX designers in addressing privacy considerations?
    \item[RQ3.] What strategies or approaches can be adopted to better incorporate privacy into UI/UX design, avoiding deceptive practices?
\end{enumerate}

To answer these questions, we conduct a three-part qualitative study, firstly grounding our work in a systematic literature review of 102 related works. The insights therein form the basis for 15 semi-structured interviews with UI/UX practitioners with varying levels of expertise. Combining the results of these studies yields a collection of 14 distinct privacy considerations in UI/UX design, as well as 14 key factors that influence these considerations in practice. Synthesizing our findings into a practical artifact for use in UI/UX design, we create a UI/UX Privacy Pattern Catalog that summarizes and illustrates eight core privacy design patterns in UI/UX, connecting these to the considerations and factors we previously uncovered. To achieve triangulation, we validate the pattern catalog in a third study consisting of a series of two interactive workshops and follow-up survey with 11 UI/UX practitioners.

Our work contributes both conceptually and practically to the study of privacy in UI/UX, as it not only systematizes the extant literature but also helps designers in practice to think more critically about privacy in the context of their work. We achieve this through the presentation and release of practically relevant artifacts, which we design to champion informed and proactive privacy-preserving UI/UX design. Concretely, we make the following contributions:

\begin{enumerate}
    \item Through the review of relevant literature and the incorporation of practitioner perspectives, we systematize privacy considerations in UI/UX design.
    \item We supplement previous work on \textit{dark patterns} in UI/UX by focusing specifically on the privacy angle and uncovering practical factors that influence privacy in UI/UX design.
    \item We release an expert-validated UI/UX Privacy Pattern Catalog as both a PDF report and a Figma design library, serving as tools for immediate use in practice to promote privacy-preserving UI/UX design.
\end{enumerate}

\section{Related Work}
\subsection{Privacy as a UI/UX Problem}
Recent research at the intersection of HCI and UI/UX has brought to light so-called \textit{dark patterns} in the design of user interfaces on the web \cite{mathur2019,parrilli2021building}. Although a unified definition of dark patterns has yet to be reached, Mathur et al. \cite{mathur2021} point to five overarching categories of dark patterns: nagging, obstruction, sneaking, interface interference, and forced action. Furthermore, Mathur et al. \cite{mathur2021} enumerate distinct ways in which dark patterns affect users, including subverting user intent \cite{brignull2018, bosch2016, westin2019, waldman2020, day2020} or user preferences \cite{bosch2016, mathur2019, luguri2021shining, cnil2020, warner2019}, \say{tricking} users \cite{bosch2016, westin2019, maier2020}, and undermining user autonomy \cite{warner2019}. 

Connecting dark patterns directly to privacy implications, B{\"o}sch et al. \cite{bosch2016} highlight how dark patterns exploit user psychology, and they formulate \say{dark strategies} that encapsulate the ill intentions behind these patterns. They also provide examples of privacy dark patterns in the wild, such as hidden legalese and forced registration. Similar examples can be found in \cite{parrilli2021building}. The presentations of these patterns, however, are focused on the effect on the user and the exploitation of psychology \cite{Roffarello2023}, and do not consider the designer perspective. 

In contrast to the dark patterns that have pervaded the internet, other works propose \textit{privacy design strategies} as guiding principles for responsible UI/UX design. Hoepman \cite{hoepman2014} defines privacy design strategies that provide a complete classification of design patterns and their corresponding privacy-enhancing technologies as a starting point for assessing the privacy impact of existing information systems. Other work \cite{cavoukian2012privacy} points to the key considerations of context, awareness, discoverability, and comprehension in UI/UX design, albeit without a great deal of concrete recommendations or generalizable tools.

Zhang-Kennedy et al. \cite{zhangkennedy2024} document and classify types of malicious strategies and their negative impact on user information privacy. Most importantly, this study provides an excellent reference to the current state of recommendations and guidelines \cite{cnil2020, acm-guidelines2020}. In contrast to Zhang-Kennedy et al., who focus largely on the perceived responsibilities of designers and their response to others' actions, we emphasize the \textit{privacy considerations} that must be made (or, which may be missing) in the design process. In addition, we supplement the North American-based perspective of this work with a largely European participant pool.

Beyond the abovementioned works, other online sources feature either a more technical background of privacy patterns \cite{privacypatterns} or focus specifically on UI/UX \cite{brightpatterns, decentpatterns, projectsbyif2025}. These, though, do not share practical perspectives from UI/UX design.



\subsection{The Intersection of Usable Privacy and UI/UX}
Research in the area of Usable Privacy involves a number of open challenges \cite{fischer2024challenges}, among them incorporating privacy requirements in the facilitation of human-centered design \cite{groen2023achieving}. Such works find, though, that striving for usable privacy in user-centered design is complex, especially considering the uncertain nature of user privacy preferences \cite{fischer2024challenges}.

While recent work in usable privacy has focused on measuring and translating user expectations for privacy in UI or UX design, little to no work has been performed focusing on making privacy work for UI/UX \textit{designers} more usable. Furthermore, there is a lack of empirical reports of how privacy is handled currently by UI/UX designers, making it difficult to tailor usable solutions for practice. Thus, we argue that studying designers is as crucial as measuring user perceptions for privacy in UI/UX, as these designers are ultimately those tasked with conceptualizing the interfaces between humans and computers \cite{teuschel2023annoy}.

\section{Methodology}
We conduct our research in three stages: literature review, semi-structured interviews, and interactive workshops with a follow-up survey. With these different methodologies, we systematize and triangulate the role of privacy in UI/UX design, namely from the perspective of the academic literature and UI/UX practitioners, as well as through iterative feedback during the artifact design process. Our research design was submitted to and approved by our institution's IRB prior to the conduction of the research (reference \#2025-22-NM-BA).

\subsection{Systematic Literature Review}
We apply the Systematic Literature Review (SLR) methodology proposed by Kitchenham et al. \cite{Kitchenham2015}, with the goal of surveying the literature on privacy-related aspects in UI/UX. We limit the scope to works indexed by Google Scholar.

Through informal probing (of known literature sources), we compiled a list of key terms/phrases related to our literature search, and we crafted these into several search strings, found in Table \ref{tab:search_queries} (Appendix). We focused on creating individual queries that leveraged Boolean logic to include both some aspect of UI/UX \textit{and} privacy. To keep the number of results within reason, we limited the search queries to titles only.

The initial search yielded 423 results across all nine search queries, which were then filtered by the primary researcher based on title and abstract screening to reduce the list to 161 publications, i.e., to remove clearly non-relevant sources. This list was further narrowed to 102 publications following screening by two additional researchers and group discussions. The selection of literature was conducted using inclusion and exclusion criteria brainstormed by our group and then iteratively refined as the screening proceeded. Inclusion criteria included studies that examined privacy UI/UX patterns, deceptive and bright patterns, ethical design, best practices, and legal compliance. In contrast, exclusion criteria included research focusing on technical mechanisms, privacy design in physical spaces, pure usability studies, or ethical debates. The full list of included sources is found in Table \ref{tab:slr} (Appendix).

The selected literature sources were analyzed using open coding, where the primary researcher marked excerpts of note, particularly those that pointed to privacy considerations in UI/UX (RQ1), key factors that influence the role of privacy in UI/UX (RQ2), and strategies to promote privacy in UI/UX (RQ3). In addition to forming the basis of our resulting artifacts, the insights from the SLR also aided in the informed design and conduction of the interview study, discussed next.

\subsection{Semi-structured Interviews}
After the SLR, we conducted a series of semi-structured interviews with designers and developers of UI and UX currently working in the industry. The primary goal of these was to build off the findings from the SLR, and to gain insights into current perceptions and challenges with privacy work in UI/UX. To simplify how we reference our target research group, we refer to them more generally as UI/UX designers and developers without emphasizing the specifics of their background, experience, or current responsibilities.

\subsubsection{Creating the Interview Guide}
To design the interview guide for the semi-structured interviews, we followed the stages laid out by Kallio et al. \cite{Kallio2016}. The prerequisites for a semi-structured interview were fulfilled in our goal to gain practitioner insights on privacy in UI/UX, while allowing for flexibility and open-ended discussion. The second stage, retrieving and using previous knowledge, was accomplished via the basis formed in the SLR, which informed the following stages in the interview guide creation.

To formulate the preliminary interview guide, our research team held brainstorming sessions in which we defined the five primary focuses of our interviews, which were motivated by our findings from the SLR. After discussion and iteration, these primary focuses were formalized as the following:

\begin{itemize}
    \item Investigate the extent to which designers perceive privacy as an issue and whether they consider it part of their responsibilities.
    \item Explore ethical and privacy considerations in design practice, including personal opinions and professional experiences regarding privacy-preserving approaches.
    \item Inquire about challenges UI/UX practitioners may face, such as any specific challenge or barrier limiting the integration of privacy into UI/UX design.
    \item Learn about the extent to which organizations recognize and incorporate privacy considerations, exploring whether privacy is embedded in corporate policies or remains an overlooked aspect.
    \item Discover cases where privacy considerations are encouraged or not, to better explore the tensions between ethical responsibilities and organizational priorities.
\end{itemize}

By structuring the interview questions around these themes, we sought to garner insights to all three of our designed research questions, namely to understand (1) privacy considerations, (2) key factors in addressing them, and (3) strategies and approaches for better integration of privacy into UI/UX design. Based on previous literature \cite{Chenail2011, Turner2010}, we focused on formulating questions in the form of \textit{what, how,} \cite{Turner2010} and \textit{why} \cite{Barriball1994}. The interview guide was developed through several iterations with the entire research team. We discussed internally to ensure there were no ambiguities or inappropriate leading questions \cite{Chenail2011} and to keep questions open-ended to avoid potentially biased insights \cite{Turner2010}. 

Following the creation of the preliminary draft, we conducted two pilot interviews with colleagues in our network working in the UI/UX discipline. The main goal of these pilots was to identify remaining errors, ambiguities, and redundant questions in the interview guide. Following these pilots, no major changes were made. These interviews are included in our analysis. The final version of the interview guide can be found in Appendix \ref{sec:interview_guide}. 

\begin{table*}[ht]
\centering
\scriptsize
\caption{Interview participant demographics. ``Org. size'' refers to the size of the company at which the participant works, according to the EU Commission Recommendation 2003/361. ``Work Exp.'' denotes the self-reported number of years in the industry.}
\label{tab:interview_participants}
\resizebox{0.99\linewidth}{!}{
\begin{tabular}{c|l|c|c|l|c|c|c|c}
\hline
\textbf{ID} & \textbf{Role} & \textbf{Gender} & \textbf{Education} & \textbf{Industry} & \textbf{Org. size} & \textbf{Country} & \textbf{Work Exp.} & \textbf{Design Exp.} \\ \hline
I1 & User Experience Designer & Male & Master's & Energy & Large & Germany & 1 - 3 & Extensive \\
I2 & UI/UX Developer and DevOps Engineer & Male & Bachelor's & Information Technology & Micro & Albania & 5 - 10 & Extensive \\
I3 & UX researcher & Male & Doctorate & Automotive & Large & Germany & 5 - 10 & Expert \\
I4 & Software Developer & Male & Master's & Information Technology & Large & Albania & 5 - 10 & Extensive \\
I5 & UX Research & Female & Bachelor's & Information Technology & Micro & Germany & 1 - 3 & Moderate \\
I6 & Managing Director & Male & Bachelor's & Information Technology & Micro & Germany & 3 - 5 & Extensive \\
I7 & Consultant & Male & Master's & Information Technology & Large & Sweden & 3 - 5 & Extensive \\
I8 & Senior UX/UI Designer & Female & Master's & Fitness and Health & Large & Germany & 5 - 10 & Expert \\
I9 & UX Designer & Female & Master's & Information Technology & Large & Germany & 3 - 5 & Extensive \\
I10 & Executive Director & Male & Bachelor's & Information Technology & Micro & Germany & 10 - 20 & Expert \\
I11 & UI/UX Designer and User Tester & Female & Master's & Material Science & Small & Germany & 1 - 3 & Extensive \\
I12 & PhD Candidate & Male & Doctorate & Automotive & Large & United States & 5 - 10 & Expert \\
I13 & Senior UX Designer & Male & Master's & Information Technology & Large & United States & 10 - 20 & Expert \\
I14 & UX Design Lead & Male & Master's & Finance & Large & Germany & 10 - 20 & Expert \\
I15 & Senior Privacy Engineer & Female & Master's & Information Technology & Large & Austria & 5 - 10 & Extensive
\end{tabular}
}
\end{table*}

\begin{table*}[ht]
\centering
\scriptsize
\caption{Workshop participants. ``N/A'' denotes a field that was not filled out by the participant. The participants are separated by their corresponding participation in one of the two workshops, with the first workshop at the top and the second at the bottom.}
\label{tab:workshop_participants}
\resizebox{0.99\linewidth}{!}{
\begin{tabular}{l|c|c|l|c|c|c|c}
\hline
\textbf{Role} & \textbf{Gender} & \textbf{Education} & \textbf{Industry} & \textbf{Org. size} & \textbf{Country} & \textbf{Work exp.} & \textbf{Design exp.} \\ \hline
Innovation/Business Developer & Female & Master's & Software & Large & Germany & 1 - 3 & Moderate \\ 
UI/UX designer & Female & Master's & Chemical & Small & Germany & 1 - 3 & Extensive \\ 
N/A & Female & Bachelor's & Information Technology & N/A & India & 5 - 10 & Limited \\ 
Student & Female & Master's & Information Technology & N/A & Germany & 1 - 3 & Limited \\ \hline
Marketing manager & Female & Bachelor's & Information Technology & Medium & Belgium & 5 - 10 & Limited \\ 
Lead UX Designer & Male & Bachelor's & Information Technology & Medium & India & 3 - 5 & Extensive \\ 
Product Designer & Female & Master's & Retail & Large & Germany & 10 - 20 & Extensive \\ 
Web Designer & Female & Bachelor's & Multimedia Arts & Micro & Philippines & 1 - 3 & Extensive \\ 
Product designer & Female & Bachelor's & Information Technology & Micro & Bolivia & 1 - 3 & Limited \\ 
Student & Female & Master's & N/A & N/A & Germany & 1 - 3 & Limited \\ 
Junior Product Manager & Female & Master's & Healthcare & Medium & Germany & 3 - 5 & Moderate \\ 
\end{tabular}
}
\end{table*}

\subsubsection{Participant Recruitment}
For interview participant recruitment, we utilized a mixture of LinkedIn and in-person events. On LinkedIn, we used the terms \say{senior ui/ux designer}, \say{ui/ux designer}, \say{product designer}, and \say{frontend developer}, prioritizing the best-fitting profiles and those with greater experience, but also welcoming participants whose role involved researching or managing UI/UX design. We also approached many attendees at several local design-related events in our home city, asking them personally to participate in our interview study.

In both cases, we sent a formal invitation via email, outlining our main research goals, stating the estimated duration (one hour) and medium (online, Zoom), and providing a Calendly link for convenient scheduling. After an interview appointment was made, we would send the interview guide in advance, ensuring that the participant felt comfortable and prepared to answer our questions.

Of the 45 people contacted via LinkedIn, we successfully conducted interviews with four. In addition, we held interviews with eight practitioners who were approached at local events. Finally, three additional participants were interviewed after being referred by previous participants. Thus, we conducted 15 interviews in total, and we stopped recruiting participants after theoretical saturation, which was reached when no new codes arose from the thematic analysis being conducted in parallel (discussed in the following). The interviews were conducted in English, and they took place in the time period from January to March 2025. Interview participants were not compensated; however, we offered them to be included in our future communications about our research outcomes, including the artifacts shared in this work.

The complete demographics of our interview participants can be found in Table \ref{tab:interview_participants}, which shows diversity in roles, genders, education, industry, company size, and experience. We also categorize the \textit{design experience} of each participant, where \textit{limited} means entry-level experience or design as a hobby, \textit{moderate} indicates contributing to some design elements in projects, \textit{extensive} involves designing as a core part of their role, and \textit{expert} extends \textit{extensive} with many years of experience in this role. While we did strive to recruit as diverse a sample as possible, we acknowledge that our participant pool is slightly male-, EU-, and Master's(+)-biased.

\subsubsection{Interview Analysis}
Following each conducted interview, the audio recording was transcribed using Otter.ai, after which the transcription was manually corrected for errors. We then conducted a Thematic Content Analysis \cite{Anderson2007, Cruzes2011}, to extract and analyze the prevalent and recurring themes. 

Each transcript was analyzed in two steps. The focus of the first iteration was to examine each sentence of the interview text, where line-by-line coding of all transcripts allowed us to extract excerpts of interest. This stage was performed solely by the primary researcher. In the second stage, we conducted axial coding, where all highlighted excerpts were assigned to codes, after which codes were aggregated into overarching themes that could be observed from the interviews. This process was done collaboratively and on a weekly basis. The entire two-step analysis process was performed as soon as possible after each interview, in line with constant comparison \cite{glaser1965constant}, in order to enable insights to inform ensuing interviews. The complete codebook can be found in Figure \ref{fig:codebook} ofn the Appendix.

\subsection{Artifact Creation, Interactive Workshops, and Survey}
\label{sec:workshop_method}
To synthesize the findings of the SLR and practical insights from the interview study, we created a final artifact in the form of a UI/UX Privacy Pattern Catalog. We did so in line with Design Science Research (DSR), which promotes the creation and evaluation of artifacts meant to address real-world problems, but with a basis in theory (or, \say{knowledge base}) and evaluated by practitioners (or, the \say{environment}) \cite{hevner2007three,prat2014artifactevaluation}. The design decisions behind the catalog, as well as its structure, are discussed in Section \ref{sec:catalog}.

To evaluate our artifact iteratively in the \textit{relevance cycle} of DSR, we held a series of two workshops with UI/UX practitioners and enthusiasts, with the primary goal of introducing the catalog draft and receiving feedback on its structure and content. The workshops were planned for 1.5 hours, which included an introduction of our research and the catalog (45 minutes), and an open-ended and interactive discussion round among the participants (45 minutes). All participants were given a printed version of the catalog for review.

Following the implementation of the open-ended feedback gained during the two workshops, we invited the same participants to complete an online survey, in order to validate the implemented improvements to the catalog. Beyond background questions, the survey consisted of two sections: \textit{catalog feedback} and \textit{perceived outcomes}. The first set of questions (7) asks specifically about the content, structure, and presentation of the catalog. The second, perceived outcomes, gauges the participant's perceived usefulness and helpfulness of the catalog in practice. All of these questions contained response options on the five-point Likert scale, from \textit{strongly disagree} to \textit{strongly agree}. The full set of survey questions can be found in Appendix \ref{sec:survey_questions}.

Recruitment for the workshops was performed through shared contacts and recruitment channels of the events attended previously to recruit interview participants. We held the two separate workshops in April 2025 and July 2025, and we administered the follow-up survey in August-September 2025. The workshops were advertised via two separate calls to the abovementioned channels, and they were attended by four and seven participants, respectively, with no repeat participants. The survey was completed by all 11 participants.

\section{RQ1: What are the Key Privacy Considerations in UI/UX Design?}
This section introduces key privacy considerations in UI/UX design (in the following abbreviated as \say{PrC}), focusing on the recurring questions and concerns that arise when designing digital products and services involving personal data. The considerations arise out of our conversations with UI/UX designers and developers in the interview study, and they are supported by relevant literature from the SLR. The 14 privacy considerations are grouped under four categories, which were aggregated based on the coding process from the thematic analysis of the interview data. Each PrC is prefaced by an overarching question, which serves as a guide in considering this particular aspect of privacy in UI/UX design. The complete list of PrCs is found in Table \ref{tab:privacy_considerations}.

\subsection{Understanding the People Behind Privacy Decisions}

\begin{center}
\textbf{PrC1}: \textit{Are our privacy consent flows designed for the people who use them?}
\end{center}

\textbf{Designers should consider the diversity of users when designing for privacy.}
Designers are usually urged to \say{design for the user}, yet privacy consent flows frequently assume an idealized, digitally literate audience and neglect the diversity of real users. Standardized choices often leave people feeling misled or manipulated, with many consenting without comprehension and later regretting their decisions \cite{nouwens2020consentpopup}. Speculative design workshops similarly found that even well-intentioned teams created solutions for digitally literate, skeptical people, while struggling to account for first-time or stressed users \cite{nelissen2022}. Comparable patterns appear in studies of self-sovereign identity wallets, where oversharing stemmed not from indifference but from inadequate contextual support \cite{teuschel2023annoy}. Those least equipped to manage complex consent flows are often the most vulnerable, underscoring the need to embed privacy into personas not just by demographics or devices, but by attitudes, levels of understanding, and emotional responses. Inclusive, usable privacy begins with testing flows against this diversity, considering, for example, the anxious traveler, the multitasking parent, or the teenager downloading a game.

\begin{center}
\textbf{PrC2}: \textit{Do users understand what is happening with their data in context?}
\end{center}

\textbf{Designers cannot just show people privacy-related information -- they have to help users make sense of it.}
Designers often treat transparency as a matter of displaying the right words, but when information is buried in dense text, shown at the wrong time, or framed in unfamiliar terms, it becomes effectively meaningless. Users routinely misunderstand what data is collected, who can access it, and what risks are involved; even when clear information is available, many rely on guesswork or habits from other applications \cite{habib2022usability}. In one study, participants disclosed sensitive information largely because they trusted the app's branding rather than understanding the actual data exchange \cite{teuschel2023annoy}. These findings highlight that effective privacy design must go beyond making data practices technically visible; by framing explanations clearly, using plain language, and presenting data in familiar formats, designers can help users genuinely understand relevant data practices and feel that their choices matter.

\begin{center}
\textbf{PrC3}: \textit{Are we overwhelming users with privacy decisions or legal language?}
\end{center}

\textbf{If our privacy interface feels like reading the terms and conditions of a mortgage, there is a problem.}
Most users lack the time or energy to parse complex settings, lengthy policies, or ambiguous terms such as \say{data enhancement} or \say{personalization}, yet many applications still present privacy content in these forms. Faced with overload, users often skip, guess, or default to the quickest path, usually \say{accept all}, without understanding what they have agreed to \cite{nouwens2020consentpopup}. Research shows that interfaces frequently reinforce this tendency through dark patterns such as \textit{bad defaults} or \textit{hidden legalese stipulations} \cite{bosch2016}. To counter this, designers can simplify choices by using plain language, presenting one decision at a time, and offering details only when users request them. If a consent flow reads like a legal contract or feels like a trick, it signals that the design prioritizes compliance over comprehension and needs to change.

\begin{center}
\textbf{PrC4}: \textit{Have we tested our privacy consent flows for clarity and comprehension?}
\end{center}

\textbf{Designing for privacy is not finished when the interface looks clean; it is finished when users actually understand what is happening, which requires testing.}
Usability testing is standard in design, yet privacy is too often reduced to compliance: if text is visible and a checkbox is present, the requirement is assumed met. In practice, users frequently misinterpret prompts, overlook them, or believe they have opted out when in fact they have consented to broad data collection \cite{habib2022usability, fradkin2024ssrn}. Such errors reflect design shortcomings rather than user error and underscore the need to evaluate comprehension with the same rigor as other aspects of usability. Simple questions to users, such as \say{What happens if you press this?} or \say{Who can see this data?}, can reveal gaps by exposing mismatches between user expectations and system behavior. Such comprehension testing should begin early in prototyping and extend through usability and A/B studies, since privacy concerns often emerge only in realistic contexts \cite{wong2019}.

\subsection{Embedding Privacy into the Design Process}

\begin{center}
\textbf{PrC5}: \textit{Are privacy considerations reflected in Figma files or documentation?}
\end{center}

\textbf{If privacy is not visible during the design process, it will probably not appear in the final product.}
In many teams, privacy considerations surface only during legal review, backend workflows, or on the settings page, making meaningful change difficult. Studies show that privacy is largely absent from common design tools such as Figma files, journey maps, and design systems \cite{teuschel2023annoy}, and interviews with practitioners reveal it is rarely documented unless explicitly prompted, even when teams express concern. This omission frames privacy as outside the scope of design and the responsibility of others. Research in HCI and PbD shows, however, that designers do in fact play a critical role in shaping how privacy is represented and communicated \cite{wong2019}. Embedding privacy directly into design artifacts -- by marking consent points, highlighting data-sensitive screens, or making privacy decisions explicit in shared tools such as Figma, Miro, or Jira -- can help teams surface friction, identify dark patterns early, and align perspectives across disciplines \cite{wong2019}.

\begin{center}
\textbf{PrC6}: \textit{Have we involved legal, development, and business teams in privacy discussions?}
\end{center}

\textbf{Designing for privacy cannot be the job of a single person. It has to be built into the team's process, roles, and conversations.}
Organizations often treat privacy as the responsibility of legal or engineering teams, leaving designers to simply \say{make it usable}. This siloed approach fosters miscommunication and poor user experiences, as critical decisions are overlooked when privacy is not addressed collaboratively across roles \cite{zhangkennedy2024}. Effective practice requires shared ownership: collaborative privacy reviews, co-created design annotations, and cross-functional critique sessions help position privacy as an ongoing design concern rather than a compliance checklist \cite{nelissen2022}. Crucially, this also means creating space for dissent, enabling designers to question manipulative patterns and advocate for user-centered alternatives even when they conflict with business defaults \cite{zhangkennedy2024}. Embedding privacy into organizational culture in this way ensures it is negotiated transparently and integrated into the user experience from the outset.

\begin{center}
\textbf{PrC7}: \textit{Have we considered how privacy design affects user trust?}
\end{center}

\textbf{Trust is not a feature but the outcome of every design choice designers make.}
User trust depends on whether individuals feel respected, informed, and in control, and it is quickly eroded when interfaces create feelings of manipulation, deception, or surveillance. Research shows that trust is influenced by visual design, tone of voice, clarity of explanations, and the ease with which settings can be adjusted, while even subtle interface choices, such as hiding options, using loaded language, or delaying consent prompts, can significantly undermine user trust \cite{schaub2015design, habib2022usability, gunawan2022redress, teuschel2023annoy}. Studies further highlight the close relationship between usability and trust: transparent, accessible privacy controls foster security and satisfaction, whereas cluttered dashboards or obscure settings generate confusion and suspicion \cite{Zimmermann2014dashboards, vanGogh2017}. Ultimately, trust is not built by telling users they are safe, but by showing them -- through consistent, respectful, and comprehensible design choices -- that their privacy matters.

\subsection{Designing for Transparency and Control}

\begin{center}
\textbf{PrC8}: \textit{When and how are we asking for user consent?}
\end{center}

\textbf{Consent is not just a checkbox but also a conversation. And timing is everything.}
Consent is often requested at inopportune moments -- immediately upon landing on a page, midway through a task, or hidden within settings, using lengthy text, missing context, or ambiguous options. Poorly timed or obscured requests lead users to accept without reflection or feel manipulated \cite{habib2022usability, nouwens2020consentpopup}, while in some cases tracking begins before consent is obtained, undermining both trust and legitimacy \cite{papadogiannakis2021cookie}. Well-designed systems instead request consent before any data is collected, explain practices in clear terms, and present options that are equally easy to choose, yet many real-world interfaces still delay \say{reject} buttons or make them harder to act upon \cite{papadogiannakis2021cookie}. Making consent meaningful, therefore, requires deliberate attention to timing, framing, and the context users need to decide. 

\begin{center}
\textbf{PrC9}: \textit{Can users easily opt in or out of tracking or data sharing?}
\end{center}

\textbf{Saying \say{you have a choice} means nothing if the choice is buried, broken, or biased.}
Designing for control requires that users can revise choices easily, yet many interfaces make opting out far harder than opting in. Research identifies deceptive patterns such as tiny \say{reject} buttons, convoluted language, and multi-step opt-out processes compared to one-click acceptance, all of which undermine user autonomy \cite{fradkin2024ssrn, nouwens2020consentpopup}. Even when opt-outs exist, users may overlook them or assume the service will not work without consent \cite{prillard2024onboarding}. Experimental studies confirm that people are far more likely to consent when granular controls are hidden, delayed, or difficult to access, producing what scholars call \say{consent theater} -- an illusion of choice rather than meaningful control \cite{habib2022usability, teuschel2023annoy, bosch2016, gunawan2022redress}. A true user-centered approach requires parity, ensuring that opt-in and opt-out options are equally visible, accessible, and respected.

\begin{table*}[ht]
\centering
\small
\caption{Combined privacy consideration questions with their corresponding relevance to UI/UX design. }
\label{tab:privacy_considerations}
\resizebox{0.99\linewidth}{!}{
\begin{tabular}{l|l|l}
\hline
\textbf{Question} & \textbf{Relevance} & \textbf{Interviews} \\
\hline
Are our privacy consent flows designed for the people who use them? & UX Alignment & I5, I12, I13 \\
Do users understand what is happening with their data in context? & UX Alignment / Transparency & I9, I11 \\
Are we overwhelming users with privacy decisions or legal language? & UX Alignment / Transparency & I6, I8, I12 \\
Have we tested our privacy consent flows for clarity and user understanding? & User Transparency / Testing & I5, I11, I13 \\
Are privacy considerations reflected in Figma files or documentation? & UX Alignment / Transparency & I11, I13 \\
Have we involved legal, development, and business teams in privacy discussions? & Legal Responsibility / Collaboration & I1, I8, I12 \\
Have we considered how privacy design affects user trust? & Ethical Responsibility / Privacy Risks & I12, I10 \\
When and how are we asking for user consent? & User Consent & I6, I9, I13 \\
Can users easily opt in or out of tracking or data sharing? & User Consent / User Control & I4, I6, I9 \\
Have we explained in plain language what data we collect and why? & User Transparency & I6, I9, I11 \\
What data are we collecting and is it necessary? & Data Minimization & I5, I6, I10, I13 \\
Are we collecting sensitive data, and have we justified it? & Data Minimization / Privacy Risks & I8, I9 \\
Are we complying with legal frameworks and going beyond the bare minimum? & Legal Responsibility & I10, I12 \\
Are there any deceptive or manipulative interface patterns? & Ethical Responsibility & I4, I5, I8 \\
\end{tabular}
}
\end{table*}

\begin{center}
\textbf{PrC10}: \textit{Have we explained in plain language what data we collect and why?}
\end{center}

\textbf{People should not need a law degree to understand what is happening to their data, but that is exactly what applications ask of them too often.}
Practitioners often explain data practices in overly technical terms or hide behind legal jargon, which may meet regulatory requirements but leaves users confused, misled, or falsely reassured. This lack of clarity is widespread: privacy interfaces frequently rely on ambiguous phrases like \say{enhancing your experience} or \say{customizing content}, masking practices such as cross-site tracking and data sales, while bundling multiple data types under vague headings further obscures consent. Studies document that these are not accidental missteps but deliberate dark patterns such as \textit{ambiguous wording} and \textit{hidden legalese stipulations} that create the illusion of informed consent while concealing crucial details \cite{bosch2016}. Addressing this requires treating clarity as a design responsibility: using everyday language, placing explanations at the moment of decision-making, and providing layered notices that support comprehension and trust. If people cannot understand what they are consenting to, their consent cannot be considered informed.

\subsection{Ethical Boundaries, Legal Compliance, Data Discipline}

\begin{center}
\textbf{PrC11}: \textit{What data are we collecting and is it necessary?} 
\end{center}

\textbf{Just because data collectors can collect particular data, it does not mean they should.}
Treating data as currency often leads organizations to collect more than is necessary, whether from habit, stakeholder pressure, or the assumption that \say{analytics might be useful later}, but this carries serious ethical and legal risks. The principle of data minimization, embedded in the GDPR, requires collection to be only what is \say{adequate, relevant, and necessary} for a specified purpose, yet many interfaces still nudge users into oversharing through vague promises like \say{improving experience}, pre-ticked consent boxes, or hidden opt-outs. Such practices systematically violate minimization principles, with patterns like \textit{privacy zuckering} manipulating users into disclosing more than they intended \cite{bosch2016}. Addressing this requires intentional design: mapping what data a feature truly needs, identifying dependencies early, and offering clear explanations and real opportunities to decline. 

\begin{center}
\textbf{PrC12}: \textit{Are we collecting sensitive data, and have we justified it?}
\end{center}

\textbf{Sensitive data raises the stakes. If applications are going to ask for it, they need a really good reason and an even better design.}
Certain categories of personal information, such as location, biometrics, health records, sexual orientation, and political beliefs, are considered sensitive under the GDPR and other regulations, and collecting them creates heightened risks for users and products. Yet many interfaces blur the line between sensitive and \say{regular} data, asking for birthdates \say{to personalize content} or for location data to \say{improve services}, while obscuring their true purpose, such as targeted advertising \cite{valoggia2024}. Legally, sensitive data requires explicit consent, but legality alone does not ensure ethics: unclear wording, coercive designs, or penalizing trade-offs undermine meaningful choice. Better practice demands explicit opt-in, layered explanations, and reassurances about storage and retention, with research showing that approaches like visualizing data and offering granular controls can improve comprehension and trust \cite{nouwens2020consentpopup}. 

\begin{center}
\textbf{PrC13}: \textit{Are we complying with legal frameworks and going beyond the bare minimum?}
\end{center}

\textbf{Meeting the law is a minimum requirement, not a design goal.}
It can be tempting for design teams to treat compliance as the end goal, yet compliance on paper is often just the baseline that does not ensure a usable or trustworthy experience. Some interfaces even exploit the appearance of compliance, displaying \say{GDPR-compliant} labels while relying on grouped permissions, hidden defaults, or manipulative copy. Research shows this is not merely a matter of bad faith but also of difficulty: designers reported uncertainty about how to translate GDPR into actionable UI choices and at times justified questionable practices as simply \say{following orders} \cite{nelissen2022}. This underscores a deeper challenge: a feature may be lawful yet still manipulative if it obscures options, pressures users, or prioritizes data over dignity. Designers must therefore ask not just \say{Is this legal?} but \say{What would a truly privacy-respecting experience look like?}.

\begin{center}
\textbf{PrC14}: \textit{Are there dark patterns or deceptive designs present?}
\end{center}

\textbf{If the designed interface nudges users toward sharing data without realizing it, designers may be using a deceptive pattern, intentionally or not.}
Dark patterns are not just poor design habits but a growing concern for researchers and regulators, with research showing that more than half of popular websites use manipulative consent banners that often violate the spirit, and sometimes the letter, of the GDPR \cite{nouwens2020consentpopup}. Designers, frequently pressured to increase engagement or data collection, may unintentionally normalize these practices through familiarity and reuse \cite{zhangkennedy2024}. In response, academic and industry frameworks propose alternatives such as bright patterns, neutral choice architectures, and visual parity between opt-in and opt-out, which emphasize clear communication, balanced options, and freedom of choice without penalty \cite{zhangkennedy2024}. Yet awareness and adoption remain limited, underscoring the need for designers to recognize their responsibility: ethical UX is not only about avoiding manipulative patterns but about actively fostering transparent, user-respecting design.

\section{RQ2: What are the Key Factors in Addressing Privacy Considerations in UI/UX Design?}
Beyond the privacy considerations in UI/UX design, RQ2 explores the factors that enable designers to design in a privacy-preserving manner, and conversely, what serves as a hindrance. These factors encompass individual, ethical, cultural, legal, and technical boundaries, showing the complexities in incorporating privacy into UI/UX. In the following, we present 14 key factors under four categories, where we synthesize and describe the main findings from the interviews, with literature support. The four categories and their corresponding factors are summarized in Table \ref{tab:factors}.

\subsection{Designer Mindsets and Values}

\subsubsection{Empathy and User-Centered Intentions}
Across the interviews, many designers expressed a genuine desire to protect users by offering clarity, choice, and control. This behavior is derived as a result of human-centered values such as empathy, respect, and the instinct to avoid harm.
\begin{quote}
\small
    ``\textit{I want the user to have the best experience, but the business people want them to buy the most amount of things.}'' (I7)
\end{quote}
Designers spoke about privacy as an emotional and relational concern, not just a regulatory one. Several of them described trying to ``\textit{design the way they would want to be treated}'' (I4), or making decisions based on ''\textit{what felt fair}'' (I12). These expressions align closely with ethical design models proposed by Gunawan et al. \cite{gunawan2022redress}, who advocate for speculative and participatory approaches to surface these values before they are compromised.

This mindset also mirrors frameworks like privacy as contextual integrity \cite{nissenbaum2004}, where user expectations are shaped by social norms, not just legal rules. Designers who intuitively understand these expectations through their own experiences or through deep empathy with users are more likely to build flows that are accomplished ethically, even when exact compliance language is missing. However, as several interview participants noted, this care can erode when unsupported, especially when faced with business requirements that make this accomplishment difficult to achieve.

\begin{table}[t!]
\centering
\scriptsize
\caption{Key factors in addressing privacy considerations in UI/UX design.}
\label{tab:factors}
\resizebox{0.97\linewidth}{!}{
\begin{tabular}{|l|l|}
\hline
\textbf{Designer Mindsets and Values} & \textbf{Communication and Collaboration in Practice} \\ \hline
Empathy and User-Centered Intentions & Privacy as a Shared Responsibility \\
Ethical Friction and Rationalization & Interdisciplinary Friction and Misaligned Goals \\
Fear as a Design Driver & Communication Gaps and the Need for Shared Language \\
Literacy and Learning & Organizational Culture \\ \hline
\textbf{Systems, Tools, and Structural Supports} & \textbf{Societal and Structural Forces} \\ \hline
Design Systems & Regulation as a Moving Target for Design \\
Tools, Templates, and the Problem of Reinvention & Cultural Contexts \\
Design Maturity and the Privacy Paradox & Public Trust and Pressure \\ \hline
\end{tabular}
}
\end{table}

\subsubsection{Ethical Friction and Rationalization}
Designers are not unaware of deceptive patterns. Many interview participants described moments of ethical tension when faced with various design decisions, where they felt uncomfortable but complied anyway due to deadlines, hierarchy, or fear of slowing the team down (I6, I7, I9, I13).

In speculative design studies from the literature, even privacy-aware designers gradually rationalized questionable decisions containing ethical nuances when placed under mock stakeholder pressure, confronting statements such as \textit{this is just how the industry works} or \textit{we can add a fix later} \cite{nelissen2022}. These rationalizations were often framed as pragmatic compromises rather than harmful deliberation. As noted in a critique of deceptive patterns \cite{mulligan2020}, privacy harm is often \textit{felt} before it is \textit{formalized}, and without institutional language to express that discomfort, it often goes unspoken \cite{zhangkennedy2024}.

\begin{quote}
\small
    \textit{``It's money, it's business, is customer, it's conversion...'' (I14) }
\end{quote}

\subsubsection{Fear as a Design Driver}
In contrast to the decision-making process led by ethics, several designers described a more reactive posture by \textit{not doing anything that could get them in trouble}. Here, privacy is not framed as an experience to be improved but rather a liability to be avoided (I1, I8, I10). This perspective is more common in teams where the legal department operates as gatekeepers, and design has a minimal say in data decisions.

As there is a strong reason for relying on legal departments, fear appears in a positive form as a trigger for change. It was discussed in the interviews how some teams invested in privacy UX after a public backlash, internal whistleblowing, or regulatory audits. Others mentioned retrofitting transparency features after competitor scandals or app store policy changes.

\begin{quote}
\small
    ``\textit{It is always crucial to know what you're talking about, and there's a difference between what users might want, what users might want to feel, and then realizing what is actually appropriate privacy, and to judge whether privacy is preserved appropriately.}'' (I12)
\end{quote}

These accounts reveal that privacy design is often driven less by \textit{what feels right} and more by \textit{what feels urgent}. This urgency, however, may be driven by policies, press, or legal review, not by users. 

\subsubsection{Literacy and Learning}
Nearly every interview participant acknowledged some level of uncertainty when performing privacy-related work. Even experienced professionals hinted at ''\textit{guessing}'', ''\textit{borrowing from competitors}'', or ``\textit{just doing what we did last time}'' (I3, I7, I11).

This lack of structured training is well-documented in the literature. Pillai et al. \cite{pillai2022} found that most designers had little formal education on privacy law, deceptive patterns, or ethical reasoning frameworks, as they often come with a non-technical background. Instead, they rely on informal sources such as internal Slack threads, Medium posts, and past product examples, creating a fragmented and fragile knowledge base to address the complexity of problems.

\begin{quote}
\small
    ``\textit{I would like to have some quick and accessible information about privacy-preserving patterns, some sort of checklist, where I can check the actions that I could take, and then I can maybe dig deeper and read more about each of them (...) That would be definitely helpful!}'' (I8)
\end{quote}

The absence of privacy design systems, access to standardized components, pattern libraries, or internal guidance for addressing privacy in a user-centered way was a recurring pain point met during interviews. None of the interviewees shared any source they follow to address privacy in their design, despite their interest in such a resource.

\begin{tcolorbox}
    \textbf{Key takeaway:} Designers \textit{do} care about privacy. But this care is unfortunately not sufficient without the proper support structures and when faced with persistent pressure to ship. We find that a number of personal factors shape privacy decisions in design practices, and how these forces affect privacy in UI/UX design depends on how they are fostered.
\end{tcolorbox}

\subsection{Communication and Collaboration in Practice}

\subsubsection{Privacy as a Shared Responsibility}
We find a consistent theme related to the diffusion of privacy responsibility, where designers described deferring to legal, compliance, or development teams. This is exemplified by one interview participant: ``\textit{I just waited around to hear back from legal...}'' (I1).

Similarly, Zhang-Kennedy et al. \cite{zhangkennedy2024} found that many designers felt a moral obligation to protect users, but they were structurally excluded from privacy decisions, either because their organizations had centralized privacy within legal teams or because developers ultimately implemented the logic. Designers, on the other hand, described their role as making sure the UI does not contradict compliance but not about shaping privacy strategy. This siloed structure leads to missed opportunities for privacy to be considered more as a user experience rather than a compliance task. Designers cannot meaningfully advocate for user control if they are continuously excluded from how data is stored, processed, or logged.

\subsubsection{Interdisciplinary Friction and Misaligned Goals}
Several interview participants described privacy as a kind of \say{hot potato}, something passed between teams with no clear owner (I6, I10, I12), and as expected, the result often was ambiguity, tension, and late-stage compromises. Designers reported that legal teams, when there is one, control content and timing of consent flows, while engineering teams decide on feasibility, leaving little space for nuanced UX improvements.

These coordination issues often created design troubles. Some designers recounted how privacy was added \say{at the last minute} or through \say{legal text copy-pasted into a modal}, resulting in unfriendly user flows, attributed to antagonist strategies even when unintentional, meeting the letter but not the spirit of the regulation (I2, I5). Pillai et al. \cite{pillai2022} reinforce this, describing privacy as a site of competing priorities, where designers must constantly negotiate between business goals, user needs, and legal ambiguity. Where designers have early access to privacy requirements and regular contact with legal compliance teams, designs can be more ethically aligned.

\subsubsection{Communication Gaps and the Need for Shared Language}
Another factor lies not in disagreement but in miscommunication. Designers and legal experts often speak different languages, one rooted in interaction and usability and the other in risk and regulation. This gap leads to slowdowns, misunderstandings, and inconsistent design decisions.

\begin{quote}
\small
    ``\textit{There is a significant lack, as I see that a lot of people are answering based on experience rather than a website, a set of guidelines, a wiki, a book, or anything that would be very helpful.}'' (I5)
\end{quote}

Interview participants called for supportive tools such as pattern libraries, glossaries, and checklists to help bridge this divide. Some argued that privacy guidelines were too abstract or too technical, making it difficult to apply them meaningfully in UI design (I7, I11), and others described relying on legal contacts or a privacy-savvy developer rather than any formal process.

Studies have shown designers reporting relying on peer networks, blogs, and community examples for privacy advice, but not having any internal documentation they could rely on \cite{nelissen2022}. Thus, communication becomes reactive rather than strategic, pointing to a need for a shared vocabulary for privacy integrated into team practices and systems.

\subsubsection{Organizational Culture}
Some designers we talked to pointed to organizational culture as a key factor in addressing privacy in UI/UX design. Some worked in teams with \say{privacy office hours}, \say{cross-functional UX audits}, or champion programs that supported ethical discussion (I10, I12, I14). Others described less privacy-friendly environments, with a hesitation \say{to talk about it} unless it was legally required (I1, I8).

These experiences show that organizational alignment is essential for ethical design, and without it, even privacy-aware designers are forced to prioritize shipping over safeguarding. In contrast, when privacy is integrated into team values, design systems, and performance metrics, the likelihood and sustainability of better decisions are both increased.

\begin{tcolorbox}
    \textbf{Key takeaway:} Whether or not privacy is considered in UI/UX design largely relies on collaboration. When teams talk openly about privacy and treat it as a shared responsibility, better privacy flows emerge. If privacy is siloed or treated as ``somebody else's job'', privacy may be left behind.
\end{tcolorbox}

\subsection{Systems, Tools, and Structural Supports}

\subsubsection{Design Systems}
Despite the centrality of design systems in modern UI/UX work, privacy is often missing from the daily component libraries and style guides designers use. Interview participants reported that while their systems included components like buttons, cards, or color tokens, there was rarely guidance on how to handle privacy-sensitive flows like consent, permissions, or deletion (I2, I9, I11).This can lead to duplication, inconsistency, and friction between components.

\begin{quote}
\small
    ``\textit{If we don’t have it in the system, someone will hack it together, but then it’s not always compliant or ethical}'' (I4).
\end{quote}

Similarly, Zhang-Kennedy et al. \cite{zhangkennedy2024} find that many teams lack reusable patterns or documentation for privacy tasks, leading to ad-hoc decisions and reliance on precedent rather than principle. Others, such as Nelissen and Funk \cite{nelissen2022}, find that even design-savvy professionals struggle to translate abstract privacy goals into interface decisions without concrete assets.

\subsubsection{Tools, Templates, and the Problem of Reinvention}
A recurring pain point expressed in the interviews was the absence of structured tools, guidance, or frameworks for privacy work in UI/UX. Designers mentioned relying on scattered blog posts, Figma community files, or internal Slack threads. Some teams created their own informal \say{Q\&A privacy checklists} or built dashboards for reviewing consent flows (I5, I10).

Several interview participants noted that legal documentation was often unusable in a design context. It may list requirements (\say{give users control}) but not give guidance on how that should look. Without translation into UX terms, designers defaulted to minimal compliance. There was a shared desire for design pattern libraries tailored for privacy with examples, copy templates, usage guidelines, or fallback states.
Zhang-Kennedy et al. \cite{zhangkennedy2024} find that many teams lack centralized privacy assets and instead learn through trial-and-error or mimicry. Their findings suggest that learning through mistakes is the norm and that access to vetted, reusable materials could dramatically improve both speed and ethical quality.

\subsubsection{Design Maturity and the Privacy Paradox}
We learned that team maturity plays a key role, as some interview participants shared examples of integrated workflows, e.g., privacy red flags raised during design critiques, opt-in flows tested alongside usability tasks, and deletion processes. These teams often had a privacy lead or embedded guideline that helped address edge cases early (I6, I12, I15).

This level of maturity mirrors what frameworks like PbD envision, including proactive and user-centered privacy considerations. However, studies such as Gunawan et al. \cite{gunawan2022redress} stress that most teams still struggle to operationalize these ideas and often default to reactive, compliance-driven approaches unless given clear support. We observed this from designers from less mature organizations, where they shared accounts of pushing for changes that \say{got ignored}, trying to introduce patterns that were not approved, or deferring to engineers who \say{just shipped what was easiest} (I3, I7, I13).

\begin{tcolorbox}
    \textbf{Key takeaway:} Designers do not just need privacy values but \textit{privacy scaffolding} -- design systems, templates, checklists, and collaborative tools. When privacy is built into these systems, ethical design can become the path of least resistance.
\end{tcolorbox}

\subsection{Societal and Structural Forces}

\subsubsection{Regulation as a Moving Target for Design}
Mention of regulation was also met with fear of \say{getting it wrong} (I2, I6, I8). Many of the interview participants painted privacy regulations not as enablers but as opaque, shifting targets. Although the goal is to comply, they do not always know how to do so, especially when legal requirements conflict with usability best practices.

In previous research \cite{nelissen2022}, participants were reported to be unfamiliar with how regulations are applied to design. Some thought their work was compliant when it was not; others were overwhelmed by the ambiguity of the legal checklists provided. Such studies reinforce the challenge that aligning regulatory demands in UI/UX design poses a complex task to UI/UX designers. The root issue, as researchers have argued, is that regulations were not written for designers, and so the burden of interpretation often falls on UX teams with limited guidance \cite{wong2019}. PbD, for example, is legally mandated under GDPR, but what \textit{by design} means in practice is often unclear, resulting in checklist approaches that miss deeper ethical questions \cite{leiser2022dark}.

\begin{quote}
\small
    ``\textit{Privacy by design is not really a design discipline, but more of an engineering aspect, and then you have usable privacy, which is where a lot more interface design and visual design come into play.}'' (I12)
\end{quote}

\subsubsection{Cultural Contexts}
Many interview participants noted that \textit{expectations} about privacy vary drastically across markets, user segments, and regions. For example, users in Germany or France might expect stricter consent flows than those in the United States. At the same time, healthcare platforms must navigate vastly different norms than social media or gaming apps (I8, I9, I12, I13, I15). These differences again reflect frameworks such as contextual integrity, or the idea that privacy expectations are shaped by local norms, relationships, and values \cite{barkhuus2012}. This is echoed by I4: ``\textit{What I think is respectful might feel invasive to someone else.}'' (I4)

Designers must then balance regulatory requirements, business goals, and cultural sensitivity, often without a clear playbook. In multicultural teams, they expressed concern that their own assumptions might not map onto user expectations. Some relied on user testing or research to catch these mismatches; others admitted to guessing.

\begin{quote}
\small
    ``\textit{I know what is ethical and what isn't, but it's very easy for ethical questions to become political questions or a clash of cultures.}'' (I15)
\end{quote}

Van Gogh \cite{vanGogh2017} shows that trust is a key mitigator for this factor, where people are more willing to share data when they feel the interface is honest, culturally relevant, and emotionally aligned with their expectations.

\subsubsection{Public Trust and Pressure}
We learn that designers do not just respond to rules and norms; they also respond to current events. Several interview participants mentioned that real change in their teams happened only after a public scandal, data breach, or regulatory fine (I1, I7, I10). This reflects what Mulligan et al. \cite{mulligan2020} call \textit{the fertile dark matter of privacy}, a space where users feel harm or violation long before that harm becomes formalized in law or metrics. Designers operate in this tension, knowing users care, but they struggle to prioritize privacy without public or institutional pressure.

Particularly after impactful incidents like the Cambridge Analytica scandal, where privacy is at the forefront of discussion, users may exhibit lower trust in platforms, and development teams can in turn scramble to retroactively redesign consent flows or delete features that raise concerns. Unfortunately, these redesigns may often happen under duress, not as part of a proactive, user-driven strategy \cite{schaub2015design}. These findings raise the important point that trust is a design component, and it is built through clarity, consistency, and choice.

\begin{tcolorbox}
    \textbf{Key takeaway:} Designers work within an ecosystem shaped by laws, norms, and cultural perspectives. Even the best privacy workflows may become ineffective without clarity from regulation, alignment with local expectations, and responsiveness to public concerns. Thus, such factors must be translated into the design of the user experience. 
\end{tcolorbox}

\section{RQ3: How can Designers be Better Equipped to Incorporate Privacy into UI/UX Design?}
We introduce our approach to developing a pattern catalog that supports integrating privacy-preserving practices into UI/UX design work, as well as the catalog evaluation results. 

\subsection{Designing a UI/UX Privacy Pattern Catalog}
\label{sec:catalog}
The UI/UX Privacy Pattern Catalog is created from key findings in both the SLR and interview study, with the goal of bridging theoretical findings with practitioner needs.

In order to accomplish this goal, we center the catalog on a collection of selected \textit{pattern categories}, which represent key interfaces that involve privacy-related decisions. We ground each of these categories in the literature, mapping them to dark strategies by B{\"o}sch et al. \cite{bosch2016}, privacy strategies by Hoepman \cite{hoepman2014}, and legal privacy principles by Fritsch \cite{fritsch2017}, as well as descriptive information garnered from the literature and interviews. Most importantly, we provide tangible, visual recommendations for incorporating privacy into the design of these key elements, breaking down the pattern into three key points. An example excerpt of the catalog structure can be found in Figure \ref{fig:catalog} of the Appendix, and the published catalog can be accessed at \url{https://mediatum.ub.tum.de/1840431}.

Above all, the catalog is rooted in the idea that designers prefer actionable and visual guidelines, where the recommendations for incorporating privacy into design are clearly provided. This was made clear in the interviews, where several designers mentioned the lack of any such resource to date. To add to the usability of our pattern catalog, we also design a companion Figma library, which contains real template examples of the design recommendations introduced in the catalog. The complete public library which contains all the interactive and reusable components is hosted at \url{https://www.figma.com/design/YenBmjGmzzb87oAyABmSOL}.

\begin{figure*}[ht!]
    \centering
    \includegraphics[scale=0.48]{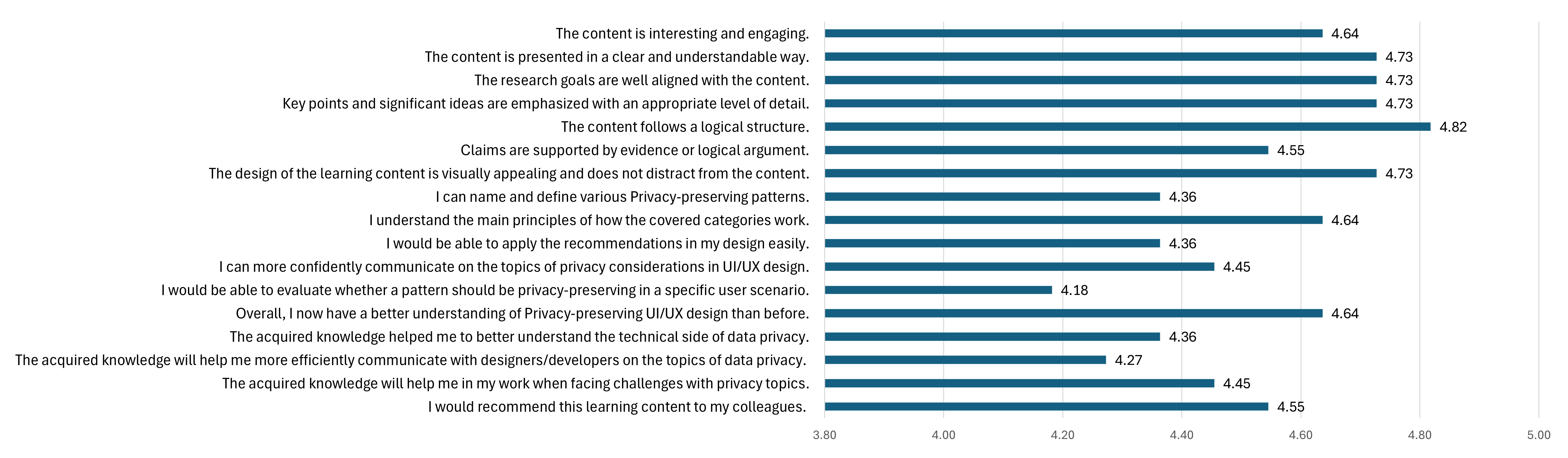}
    \caption{Aggregated survey results for our catalog validation. All values represent the average score for the 11 responses to each survey question, where \textit{strongly disagree} corresponds to a score of 1 and \textit{strongly agree} to a score of 5.}
    \label{fig:survey_results}
\end{figure*}

\subsection{Artifact Evaluation}
The goal of the two workshops, introduced in Section \ref{sec:workshop_method}, was to gain feedback on the catalog during the design process, and to refine it iteratively for further feedback and validation. 

Both workshops followed a similar structure, in which the first half was spent introducing our research findings, the in-progress catalog, and the Figma library being designed. After this, an interactive discussion round was held, where the feedback of the participants was recorded by hand. The workshops were followed by an online survey (Appendix), as also introduced in Section \ref{sec:workshop_method}.

A summary of the survey results is presented in Figure \ref{fig:survey_results}. We aggregate all responses to the Likert-type questions from the 11 survey respondents, with the corresponding questions attached. Further feedback we received during the interactive sessions is discussed in the following section.

\section{Discussion}
In the following, we reflect on the main findings of our work, drawing upon the insights from the SLR, interviews, and workshops to discuss the broader implications of our research.

\subsection{RQ1: Privacy Starts with UI}
In our review of the literature and discussions with 15 UI/UX designers, we learned of many important privacy considerations to make in design, which we summarize into 14 \textit{questions}. These questions form a self-assessment for designers and teams to perform, in order to evaluate the maturity of their privacy practices. 

The implications of these privacy considerations and their corresponding questions become clear: privacy often starts with the UI/UX. In decisions surrounding questions about the scope of data collection, the form of informing users, and compliance with relevant regulations, the implications of these decisions all become immediately and tangibly clear at the user interface, the point where such privacy considerations take form. The necessity and form of downstream privacy practice, such as in how data is stored, is directly influenced by how privacy is treated at the user interface. Thus, we reinforce the perspective that privacy begins where users first engage with a system, \textbf{the interface}. As such, privacy-preserving design does not simply protect data; it protects the relationship between the user and the system.

This places a considerable role on UI/UX designers, who traditionally are not privacy experts, but in reality do face such decisions prevalently. As such, key factors arise that lead to challenges for privacy-preserving UI/UX design, requiring tangible guidance beyond theoretical frameworks.

\subsection{RQ2/RQ3: Empowering Designers to Address Key Factors}
\begin{quote}
\small
    ``\textit{I would say 99\% of my friends are all designers. Believe me, I would say nobody cares, and maybe I am one of the few, maybe the only one, thinking about privacy and security. And so I would say not to disappoint you, but I think designers are more in leading how we, users, engage with the system, and (...) aesthetics about the system, and not really about the technical issue [of privacy].}'' (I3)
\end{quote}

Perspectives like the above suggest a possible disconnect between the idea that designers are the frontline defense for user privacy, and what is actually the predominant focus and mindset of UI/UX designers in practice today. Through the series of 14 factors influencing the consideration of privacy in UI/UX, we learn that there are many forces that make privacy-preserving UI/UX attractive in theory, but quite difficult to realize in practice. This is met with pessimistic, yet honest practitioner perceptions such as the above, which point to sparse awareness and willingness to address privacy, but perhaps with lacking general support and guidance to do so.

Therefore, as much as there is momentum around privacy-preserving UI/UX design, there is also a reality that is harder to ignore: many designers simply are not thinking about it. This is a clear call to researchers to shift emphasis to the UI/UX field, investigating the pain points and opportunities in this discipline, and also increasing work on creating usable guidelines for designers to implement in their daily work.

In the two workshops conducted with UI/UX designers, developers, and enthusiasts, we learned of a number of desirable characteristics in research artifacts that not only guided us in refining the pattern catalog, but also shed light on concrete points for better equipping designers to handle privacy in UI/UX. Above all, we receive positive feedback based on the fact that we provided concrete examples in an engaging manner, demonstrating the importance of clear visual examples for designers. Nevertheless, we also received feedback expressing concerns about the lack of connection to legal regulations, a concern we partially mitigate via the inclusion of corresponding legal privacy principles (Figure \ref{fig:catalog}). Additionally, participants involved in development pointed to potential incompatibility with other standards in UI/UX, an aspect that we acknowledge the catalog does not fully address.

The promise of the UI/UX Privacy Pattern Catalog is demonstrated by the results of Figure \ref{fig:survey_results}, which show generally high agreement with the presented structure, content, and usefulness of the catalog. In this, we make a case for the value in the dissemination of research artifacts that can be readily used in practice, adding to the usability of such research. We plan to disseminate the catalog and the companion Figma library to a wider audience following publication.

\section{Conclusion}
We investigate the nature of privacy in UI/UX design, conducting a multi-stage study that triangulates academic literature, practitioner experiences, and qualitative and quantitative feedback on our research artifact. We systematize 14 considerations that must be made for the integration of a privacy mindset into UI/UX design, and uncover 14 key factors that affect privacy in UI/UX. The insights from both the SLR and interviews guide the creation of our \textit{UI/UX Privacy Pattern Catalog}, which was iteratively designed over two interactive workshops with UI/UX designers, and subsequently validated via an online survey. We find that our catalog is perceived to be a useful resource to designers, and along with our collection of privacy considerations and factors, we provide concrete guidance to promote privacy-preserving UI/UX design.

\paragraph{Limitations}
We acknowledge the main limitations of our study, firstly, in that our theoretical foundation (SLR) is based solely on a title-only search from Google Scholar. While Scholar provides a comprehensive indexing of numerous sources, some relevant papers may have been missed.

Our results are largely biased to the European mindset, with relatively few voices from outside of this region. In addition, the open-ended nature of the conducted workshops precluded controlled questioning and steered conversations, resulting in limited ability for comprehensive insights or rigorous analysis (as the workshops were not recorded). 

In addition to the above, we do not distinguish between UI and UX, and furthermore, we did not analyze the interview data in the specific work context of each interview participant. While this increases the generalizability of our findings, it does not take into account domain- or industry-specific factors which may further affect privacy considerations (for example, in more sensitive industries such as healthcare or finance). 

\paragraph{Future Work}
In addition to addressing the limitations above, future work should seek to investigate further the degree to which a current gap exists between privacy-preserving UI/UX theory, and actual awareness and implementation in practice. To aid in this, we seek to refine the \textit{UI/UX Privacy Pattern Catalog} and the companion Figma library, with the goal of making these living artifacts. In all of these cases, we see it as important for future research to continue to include practitioner voices in the research, design, and validation of practice-oriented privacy-preserving UI/UX guidance.





%


\newpage
\bibliographystyle{IEEEtran}
\bibliography{bibliography}

@book{Kitchenham2015,
  author    = {B. A. Kitchenham and D. Budgen and P. Brereton},
  title     = {Evidence-based software engineering and systematic reviews},
  volume    = {4},
  publisher = {CRC Press},
  year      = {2015}
}

@article{Kallio2016,
  author    = {H. Kallio and A.-M. Pietilä and M. Johnson and M. Kangasniemi},
  title     = {Systematic methodological review: developing a framework for a qualitative semi-structured interview guide},
  journal   = {Journal of Advanced Nursing},
  volume    = {72},
  number    = {12},
  pages     = {2954--2965},
  year      = {2016},
  doi       = {10.1111/jan.13031}
}

@article{Chenail2011,
  author    = {R. J. Chenail},
  title     = {Interviewing the investigator: Strategies for addressing instrumentation and researcher bias concerns in qualitative research},
  journal   = {Qualitative Report},
  volume    = {16},
  number    = {1},
  pages     = {255--262},
  year      = {2011},
  doi       = {10.46743/2160-3715/2011.1051}
}

@article{Turner2010,
  author    = {{Turner, D. W. III}},
  title     = {Qualitative interview design: A practical guide for novice investigators},
  journal   = {The Qualitative Report},
  volume    = {15},
  number    = {3},
  pages     = {754},
  year      = {2010},
  doi       = {10.46743/2160-3715/2010.1178}
}

@article{Barriball1994,
  author    = {K. L. Barriball and A. While},
  title     = {Collecting data using a semi-structured interview: a discussion paper},
  journal   = {Journal of Advanced Nursing-Institutional Subscription},
  volume    = {19},
  number    = {2},
  pages     = {328--335},
  year      = {1994},
  doi       = {10.1111/j.1365-2648.1994.tb01088.x}
}

@article{zhangkennedy2024,
  title={Navigating the Gray: Design Practitioners' Perceptions Toward the Implementation of Privacy Dark Patterns},
  author={Zhang-Kennedy, Leanne and Keleher, Michael and Valiquette, Mathieu},
  journal={Proceedings of the ACM on Human-Computer Interaction},
  volume={8},
  number={CSCW1},
  pages={1--26},
  year={2024},
  publisher={ACM},
  doi={10.1145/3637374}
}

@article{Anderson2007,
  author    = {R. Anderson},
  title     = {Thematic content analysis ({TCA}): Descriptive presentation of qualitative data},
  journal   = {Descriptive Presentation of Qualitative Data},
  volume    = {3},
  pages     = {1--4},
  year      = {2007}
}

@inproceedings{Cruzes2011,
  author    = {D. S. Cruzes and T. Dyb{\aa}},
  title     = {Recommended steps for thematic synthesis in software engineering},
  booktitle = {2011 International Symposium on Empirical Software Engineering and Measurement},
  pages     = {275--284},
  year      = {2011},
  publisher = {IEEE},
  doi       = {10.1109/ESEM.2011.36} 
}

@inproceedings{pillai2022,
author = {Pillai, Ajit G. and Sachathep, Thida and Ahmadpour, Naseem},
title = {Exploring the experience of ethical tensions and the role of community in {UX} practice},
year = {2022},
isbn = {9781450396998},
publisher = {Association for Computing Machinery},
address = {New York, NY, USA},
url = {https://doi.org/10.1145/3546155.3546683},
doi = {10.1145/3546155.3546683},
booktitle = {Nordic Human-Computer Interaction Conference},
articleno = {60},
numpages = {13},
location = {Aarhus, Denmark},
series = {NordiCHI '22}
}

@article{teuschel2023don,
  title={’Don’t Annoy Me With Privacy Decisions!’—Designing Privacy-Preserving User Interfaces for SSI Wallets on Smartphones},
  author={Teuschel, Moritz and P{\"o}hn, Daniela and Grabatin, Michael and Dietz, Felix and Hommel, Wolfgang and Alt, Florian},
  journal={IEEE Access},
  volume={11},
  pages={131814--131835},
  year={2023},
  publisher={IEEE}
}

@article{nelissen2022rationalizing,
  title={Rationalizing dark patterns: Examining the process of designing privacy UX through speculative enactments},
  author={Nelissen, LGM and Funk, Mathias},
  journal={International Journal of Design},
  volume={16},
  number={1},
  pages={75--92},
  year={2022},
  publisher={National Taiwan University of Science and Technology}
}

@inproceedings{sanchez2023ethical,
  title={Ethical tensions in UX design practice: exploring the fine line between persuasion and manipulation in online interfaces},
  author={S{\'a}nchez Chamorro, Lorena and Bongard-Blanchy, Kerstin and Koenig, Vincent},
  booktitle={Proceedings of the 2023 ACM designing interactive systems conference},
  pages={2408--2422},
  year={2023}
}

@article{nissenbaum2004,
  title={Privacy as contextual integrity},
  author={Nissenbaum, Helen},
  journal={Washington Law Review},
  volume={79},
  pages={119--157},
  year={2004}
}

@article{nelissen2022,
  title={Rationalizing dark patterns: Examining the process of designing privacy {UX} through speculative enactments},
  author={Nelissen, L. G. M. and Funk, M.},
  journal={International Journal of Design},
  volume={16},
  number={1},
  pages={75--92},
  year={2022}
}

@inproceedings{mathur2021,
author = {Mathur, Arunesh and Kshirsagar, Mihir and Mayer, Jonathan},
title = {What Makes a Dark Pattern... Dark? {D}esign Attributes, Normative Considerations, and Measurement Methods},
year = {2021},
isbn = {9781450380966},
publisher = {Association for Computing Machinery},
address = {New York, NY, USA},
url = {https://doi.org/10.1145/3411764.3445610},
doi = {10.1145/3411764.3445610},
booktitle = {Proceedings of the 2021 CHI Conference on Human Factors in Computing Systems},
articleno = {360},
numpages = {18},
location = {Yokohama, Japan},
series = {CHI '21}
}

@inproceedings{wong2019,
author = {Wong, Richmond Y. and Mulligan, Deirdre K.},
title = {Bringing Design to the Privacy Table: Broadening “Design” in “Privacy by Design” Through the Lens of {HCI}},
year = {2019},
isbn = {9781450359702},
publisher = {Association for Computing Machinery},
address = {New York, NY, USA},
url = {https://doi.org/10.1145/3290605.3300492},
doi = {10.1145/3290605.3300492},
booktitle = {Proceedings of the 2019 CHI Conference on Human Factors in Computing Systems},
pages = {1–17},
numpages = {17},
location = {Glasgow, Scotland Uk},
series = {CHI '19}
}

@article{waldman2020,
  title={Cognitive biases, dark patterns, and the ‘privacy paradox’},
  author={Waldman, Ari Ezra},
  journal={Current Opinion in Psychology},
  volume={31},
  pages={105--109},
  year={2020},
  doi={10.1016/j.copsyc.2019.08.025},
  note={Privacy and Disclosure, Online and in Social Interactions}
}

@article{teuschel2023annoy,
  author    = {Teuschel, Markus and Pöhn, Dominik and Grabatin, Maximilian and Dietz, Florian and Hommel, Wolfgang and Alt, Florian},
  title     = {``{D}on’t Annoy Me With Privacy Decisions!'' -- Designing Privacy-Preserving User Interfaces for {SSI} Wallets on Smartphones},
  journal   = {IEEE Access},
  volume    = {11},
  pages     = {131814--131835},
  year      = {2023},
  publisher = {IEEE},
  doi       = {10.1109/ACCESS.2023.3335027}
}

@article{bosch2016,
  title={Tales from the Dark Side: Privacy Dark Strategies and Privacy Dark Patterns},
  author={B{\"o}sch, Christoph and Erb, Benjamin and Kargl, Frank and Kopp, Henning and Pfattheicher, Stefan},
  journal={Proceedings on Privacy Enhancing Technologies},
  volume={2016},
  number={4},
  pages={237--254},
  year={2016},
  publisher={De Gruyter},
  doi={10.1515/popets-2016-0038}
}

@inproceedings{hoepman2014,
  title={Privacy design strategies},
  author={Hoepman, Jaap-Henk},
  booktitle={IFIP International Information Security Conference},
  pages={446--459},
  year={2014},
  publisher={Springer, Berlin, Heidelberg}
}

@misc{brignull2018,
  author = {Brignull, Harry},
  title = {Dark Patterns},
  year = {2018},
  howpublished = {\url{https://darkpatterns.org/} (now \url{https://www.deceptive.design/})}
}

@book{lewis2014,
  author    = {Lewis, Chris},
  title     = {Irresistible Apps: Motivational Design Patterns for Apps, Games, and Web-based Communities},
  year      = {2014},
  edition   = {1st},
  publisher = {Apress},
  address   = {Berkeley, CA, USA}
}

@article{day2020,
  author = {Day, Gregory and Stemler, Abbey},
  title = {Are Dark Patterns Anticompetitive?},
  journal = {Alabama Law Review},
  volume = {72},
  pages = {1--45},
  year = {2020},
  doi = {10.2139/ssrn.3468321}
}

@inproceedings{westin2019,
  title={Opt out of Privacy or “Go Home”: Understanding Reluctant Privacy Behaviours through the {FoMO}-Centric Design Paradigm},
  author={Westin, Fiona and Chiasson, Sonia},
  booktitle={Proceedings of the New Security Paradigms Workshop (NSPW '19)},
  pages={57--67},
  year={2019},
  publisher={Association for Computing Machinery},
  address={San Carlos, Costa Rica},
  doi={10.1145/3368860.3368865}
}

@article{mathur2019,
  title={Dark Patterns at Scale: Findings from a Crawl of 11K Shopping Websites},
  author={Mathur, Arunesh and Acar, Gunes and Friedman, Michael J and Lucherini, Elena and Mayer, Jonathan and Chetty, Marshini and Narayanan, Arvind},
  journal={Proceedings of the ACM on Human-Computer Interaction},
  volume={3},
  number={CSCW},
  pages={1--32},
  year={2019},
  publisher={ACM},
  doi={10.1145/3359183}
}

@article{luguri2021shining,
  title={Shining a light on dark patterns},
  author={Luguri, Jamie and Strahilevitz, Lior Jacob},
  journal={Journal of Legal Analysis},
  volume={13},
  number={1},
  pages={43--109},
  year={2021},
  publisher={Oxford University Press}
}

@article{cnil2020,
  title={Shaping choices in the digital world. From dark patterns to data protection: the influence of ux/ui design on user empowerment},
  author={Chatellier, R{\'e}gis and Delcroix, Geoffrey and Hary, Estelle and Girard-Chanudet, Camille and Leroux, Faget Pauline Marie and Chapelle, Stephanie},
  journal={6. IP Repoerts: Innovation and Foresight},
  year={2019},
  publisher={CNIL Paris}
}

@misc{california2020,
  author = {{California Secretary of State}},
  title = {Qualified Statewide Ballot Measures},
  year = {2020}
}

@misc{warner2019,
  author = {Warner, Mark and Fischer, Debra},
  title = {Senators Introduce Bipartisan Legislation to Ban Manipulative “Dark Patterns”},
  year = {2019}
}

@article{maier2020,
  title={Dark Design Patterns: An End-User Perspective},
  author={Maier, Maximilian and Harr, Rikard},
  journal={Human Technology},
  volume={16},
  number={2},
  pages={170--199},
  year={2020}
}

@misc{acm-guidelines2020,
  author = {{Netherlands Authority for Consumers \& Markets}},
  title  = {Guidelines: Protection of the Online Consumer – Boundaries of Online Persuasion},
  year   = {2020},
}

@inproceedings{gray2019ethical,
  title={Ethical mediation in UX practice},
  author={Gray, Colin M and Chivukula, Shruthi Sai},
  booktitle={Proceedings of the 2019 CHI Conference on Human Factors in Computing Systems},
  pages={1--11},
  year={2019},
  organization={ACM},
  doi={10.1145/3290605.330040}
}

@misc{privacypatterns,
  author       = {{Privacy Patterns}},
  title        = {PrivacyPatterns.org: A Collection of Privacy Design Patterns},
  year         = {n.d.},
  howpublished = {\url{https://privacypatterns.org}}
}

@ONLINE{brightpatterns,
  author = {Sandhaus, Hauke},
  title = {Brightpatterns.org: A Front Page to Define the Term and Collect Examples},
  year = {2023},
  url = {https://www.brightpatterns.org}
}

@misc{decentpatterns,
  author       = {{Decent Patterns}},
  title        = {Design Patterns for Ethical Tech},
  year         = {n.d.},
  howpublished = {\url{https://decentpatterns.xyz/library/}}
}

@misc{projectsbyif2025,
  author       = {{Projects by IF}},
  title        = {Catalogue of Digital Rights},
  year         = {n.d.},
  howpublished = {\url{https://catalogue.projectsbyif.com}}
}

@inproceedings{nouwens2020consentpopup,
  title={Dark patterns after the {GDPR}: Scraping consent pop-ups and demonstrating their influence},
  author={Nouwens, Midas and Liccardi, Ilaria and Veale, Michael and Karger, David and Kagal, Lalana},
  booktitle={Proceedings of the 2020 CHI Conference on Human Factors in Computing Systems},
  pages={1--13},
  year={2020},
  organization={ACM},
  doi={10.1145/3313831.3376321}
}

@inproceedings{fritsch2017,
  title={Privacy dark patterns in identity management},
  author={Fritsch, Lorenz},
  booktitle={Open Identity Summit 2017},
  pages={93--104},
  year={2017},
  publisher={Gesellschaft f{\"u}r Informatik},
  address={Bonn, Germany}
}

@inproceedings{caragay2024,
  title={Beyond Dark Patterns: A Concept-Based Framework for Ethical Software Design},
  author={Caragay, Evan and Xiong, Katherine and Zong, Jonathan and Jackson, Daniel},
  booktitle={Proceedings of the CHI Conference on Human Factors in Computing Systems (CHI '24)},
  pages={1--16},
  year={2024},
  address={Honolulu, HI, USA},
  publisher={ACM},
  doi={10.1145/3613904.3642781},
  url={https://doi.org/10.1145/3613904.3642781}
}

@inproceedings{valoggia2024,
  author    = {Valoggia, Paolo and Sergeeva, Anastasia and Rossi, Andrea and Botes, Willem M.},
  title     = {Learning from the Dark Side About How (Not) to Engineer Privacy: Analysis of Dark Patterns Taxonomies from an ISO 29100 Perspective},
  booktitle = {Proceedings of the 10th International Conference on Information Systems Security and Privacy (ICISSP)},
  year      = {2024},
  publisher = {SCITEPRESS - Science and Technology Publications},
  url       = {https://doi.org/10.5220/0012420000003648}
}

@article{prat2014artifactevaluation,
  title={Artifact evaluation in information systems design-science research–a holistic view},
  author={Prat, N. and Comyn-Wattiau, I. and Akoka, J.},
  journal={Journal of the Association for Information Systems},
  volume={15},
  number={3},
  pages={125--149},
  year={2014}
}

@inproceedings{Roffarello2023,
author = {Monge Roffarello, Alberto and Lukoff, Kai and De Russis, Luigi},
title = {Defining and Identifying Attention Capture Deceptive Designs in Digital Interfaces},
year = {2023},
isbn = {9781450394215},
publisher = {Association for Computing Machinery},
address = {New York, NY, USA},
url = {https://doi.org/10.1145/3544548.3580729},
doi = {10.1145/3544548.3580729},
booktitle = {Proceedings of the 2023 CHI Conference on Human Factors in Computing Systems},
articleno = {194},
numpages = {19},
location = {Hamburg, Germany},
series = {CHI '23}
}

@inproceedings{gray2018uxside,
  title={The dark (patterns) side of {UX} design},
  author={Gray, Colin M. and Kou, Yubo and Battles, Beth and Hoggatt, Joseph and Toombs, Austin L.},
  booktitle={Proceedings of the 2018 CHI Conference on Human Factors in Computing Systems},
  pages={1--14},
  year={2018},
  month={April},
  organization={ACM},
  doi={10.1145/3173574.3174108}
}

@inproceedings{schaub2015design,
author = {Schaub, Florian and Balebako, Rebecca and Durity, Adam L. and Cranor, Lorrie Faith},
title = {A design space for effective privacy notices},
year = {2015},
isbn = {9781931971249},
publisher = {USENIX Association},
address = {USA},
booktitle = {Proceedings of the Eleventh USENIX Conference on Usable Privacy and Security},
pages = {1–17},
numpages = {17},
location = {Ottawa, Canada},
series = {SOUPS '15}
}

@article{mulligan2020,
  author    = {Mulligan, Deirdre K. and Regan, Priscilla M. and King, Jennifer},
  title     = {The Fertile Dark Matter of Privacy Takes on the Dark Patterns of Surveillance},
  journal   = {Journal of Consumer Psychology},
  year      = {2020},
  volume    = {30},
  number    = {4},
  pages     = {767--773},
  doi       = {10.1002/jcpy.1190}
}

@inproceedings{barkhuus2012,
  author    = {Larsen Barkhuus},
  title     = {The mismeasurement of privacy: using contextual integrity to reconsider privacy in {HCI}},
  booktitle = {Proceedings of the SIGCHI Conference on Human Factors in Computing Systems},
  pages     = {367--376},
  year      = {2012},
  publisher = {ACM},
  address   = {Austin, Texas, USA},
  doi       = {10.1145/2207676.2207727}
}

@inproceedings{habib2021toggles,
  title={Toggles, dollar signs, and triangles: How to (in) effectively convey privacy choices with icons and link texts},
  author={Habib, Hana and Zou, Yixin and Yao, Yijun and Acquisti, Alessandro and Cranor, Lorrie and Reidenberg, Joel and Schaub, Florian},
  booktitle={Proceedings of the 2021 CHI Conference on Human Factors in Computing Systems},
  pages={1--25},
  year={2021},
  month={May},
  organization={ACM},
  doi={10.1145/3411764.3445387}
}

@inproceedings{habib2022usability,
  author    = {Habib, Hana and Cranor, Lorrie Faith},
  title     = {Evaluating the Usability of Privacy Choice Mechanisms},
  booktitle = {Eighteenth Symposium on Usable Privacy and Security (SOUPS 2022)},
  pages     = {273--289},
  year      = {2022},
  publisher = {USENIX Association},
  address   = {Boston, MA, USA}
}

@misc{fradkin2024ssrn,
  author    = {Fradkin, Andrey and Farronato, Chiara and Lin, Tao},
  title     = {Data Sharing and Website Competition: The Role of Dark Patterns},
  year      = {2024},
  note      = {Available at SSRN: https://ssrn.com/abstract=4920040},
  url       = {https://ssrn.com/abstract=4920040}
}

@inproceedings{guida2021,
  title={Privacy policies between perception and learning through legal design: Ideas for an educational chatbot combining rights' awareness, optimized user experience and training efficacy},
  author={Guida, Silvia},
  booktitle={PSYCHOBIT Conference Proceedings},
  year={2021}
}

@inproceedings{ghazinour2016ppvm,
  title={A usability study on the privacy policy visualization model},
  author={Ghazinour, Kambiz and Albalawi, Talal},
  booktitle={2016 IEEE 14th Intl Conf on Dependable, Autonomic and Secure Computing, 14th Intl Conf on Pervasive Intelligence and Computing, 2nd Intl Conf on Big Data Intelligence and Computing and Cyber Science and Technology Congress (DASC/PiCom/DataCom/CyberSciTech)},
  pages={578--585},
  year={2016},
  publisher={IEEE},
  month={August},
  doi={10.1109/DASC-PICom-DataCom-CyberSciTec.2016.109}
}

@article{milne2004strategies,
  title={Strategies for reducing online privacy risks: Why consumers read (or don’t read) online privacy notices},
  author={Milne, George R. and Culnan, Mary J.},
  journal={Journal of Interactive Marketing},
  volume={18},
  number={3},
  pages={15--29},
  year={2004},
  publisher={Wiley},
  doi={10.1002/dir.20009}
}

@inproceedings{phokela2024location,
  title={Embedding Privacy First Human Centric in User Experience for Mobile Applications},
  author={Phokela, Kanchanjot Kaur and Singi, Kapil and Kaulgud, Vikrant},
  booktitle={Proceedings of the 17th Innovations in Software Engineering Conference (ISEC)},
  year={2024},
  organization={ACM},
  doi={10.1145/3641399.3641414}
}

@inproceedings{luger2013consent,
  title={Consent for all: Revealing the hidden complexity of terms and conditions},
  author={Luger, Ewa and Moran, Sarah and Rodden, Tom},
  booktitle={Proceedings of the SIGCHI Conference on Human Factors in Computing Systems},
  pages={2687--2696},
  year={2013},
  publisher={ACM},
  doi={10.1145/2470654.2481371}
}

@inproceedings{abraham2022implications,
  author    = {Abraham, Mohammadhadi and Saeghe, Pooyan and McGill, Matthew and Khamis, Mohamed},
  title     = {Implications of XR on Privacy, Security and Behaviour: Insights from Experts},
  booktitle = {Proceedings of the Nordic Human-Computer Interaction Conference (NordiCHI)},
  year      = {2022},
  publisher = {Association for Computing Machinery},
  doi       = {10.1145/3546155.3547306}
}

@inproceedings{prillard2024onboarding,
  author    = {Prillard, Ophelia and Boletsis, Costas and Tokas, Shukun},
  title     = {Ethical Design for Data Privacy and User Privacy Awareness in the Metaverse},
  booktitle = {Proceedings of the 10th International Conference on Information Systems Security and Privacy (ICISSP)},
  year      = {2024},
  publisher = {SCITEPRESS},
  url       = {https://doi.org/10.5220/0012412200003644}
}

@inproceedings{gunawan2022redress,
  author    = {Gunawan, Jessy and Santos, Cristina and Kamara, Irene},
  title     = {Redress for Dark Patterns Privacy Harms? {A} Case Study on Consent Interactions},
  booktitle = {Proceedings of the ACM Conference on Computer Science and Law (CSLAW)},
  year      = {2022},
  pages     = {181--194},
  doi       = {10.1145/3511265.3550444}
}

@inproceedings{Zimmermann2014dashboards,
  author       = {Christian Zimmermann and Rafael Accorsi and G{\"u}nter M{\"u}ller},
  title        = {Privacy Dashboards: Reconciling Data-Driven Business Models and Privacy},
  booktitle    = {2014 Ninth International Conference on Availability, Reliability and Security (ARES)},
  year         = {2014},
  pages        = {152--157},
  publisher    = {IEEE},
  doi          = {10.1109/ARES.2014.26}
}

@inproceedings{papadogiannakis2021cookie,
  title={User Tracking in the Post-cookie Era: How Websites Bypass {GDPR} Consent to Track Users},
  author={Papadogiannakis, Emmanouil and Papadopoulos, Panagiotis and Kourtellis, Nicolas and Markatos, Evangelos P},
  booktitle={Proceedings of the Web Conference 2021},
  year={2021},
  doi={10.1145/3442381.3450056}
}

@incollection{leiser2022dark,
  author    = {Leiser, Martin R.},
  title     = {`{D}ark patterns': the case for regulatory pluralism between the European Union's consumer and data protection regimes},
  booktitle = {Research Handbooks in European Law},
  editor    = {Kosta, Eleni and Leenes, Ronald and Kamara, Irene},
  pages     = {240--269},
  publisher = {Edward Elgar Publishing Ltd.},
  year      = {2022},
  address   = {Cheltenham},
  doi       = {10.4337/9781800371682.00019}
}

@mastersthesis{vanGogh2017,
  author       = {Stef van Gogh},
  title        = {Privacy Dashboards: Control and Understanding of Data through Usability and User Experience},
  school       = {University of Twente},
  year         = {2017},
  type         = {Master's thesis},
  url          = {https://essay.utwente.nl/73657/}
}

@article{cavoukian2012privacy,
  title={Privacy by Design and User Interfaces},
  author={Cavoukian, Ann},
  journal={Information and Privacy Commissioner, Ontario, Canada},
  year={2012}
}

@article{mcclain2023americans,
  title={How {A}mericans view data privacy},
  author={McClain, Colleen and Faverio, Michelle and Anderson, Monica and Park, Eugenie},
  journal={Pew Research Center},
  year={2023}
}

@inproceedings{bongard2021definitely,
  title={``{I} am Definitely Manipulated, Even When {I} am Aware of it. {I}t’s Ridiculous'' - Dark Patterns from the End-User Perspective},
  author={Bongard-Blanchy, Kerstin and Rossi, Arianna and Rivas, Salvador and Doublet, Sophie and Koenig, Vincent and Lenzini, Gabriele},
  booktitle={Proceedings of the 2021 ACM Designing Interactive Systems Conference},
  pages={763--776},
  year={2021},
  doi={10.1145/3461778.3462086}
}

@article{gray2021end,
  title={End user accounts of dark patterns as felt manipulation},
  author={Gray, Colin M and Chen, Jingle and Chivukula, Shruthi Sai and Qu, Liyang},
  journal={Proceedings of the ACM on Human-Computer Interaction},
  volume={5},
  number={CSCW2},
  pages={1--25},
  year={2021},
  publisher={ACM New York, NY, USA},
  doi={10.1145/3479516}
}

@inproceedings{parrilli2021building,
  title={Building a privacy oriented {UI} and {UX} design: an introduction to its foundations and potential developments},
  author={Parrilli, Davide M and Hern{\'a}ndez-Ram{\'\i}rez, Rodrigo},
  booktitle={International Conference on Design and Digital Communication},
  pages={16--30},
  year={2021},
  organization={Springer},
  doi={10.1007/978-3-030-89735-2_2}
}

@incollection{fischer2024challenges,
  title={Challenges of Usable Privacy},
  author={Fischer-H{\"u}bner, Simone and Karegar, Farzaneh},
  booktitle={The Curious Case of Usable Privacy: Challenges, Solutions, and Prospects},
  pages={103--131},
  year={2024},
  publisher={Springer},
  doi={10.1007/978-3-031-54158-2_4}
}

@incollection{groen2023achieving,
  title={Achieving usable security and privacy through Human-Centered Design},
  author={Groen, Eduard C and Feth, Denis and Polst, Svenja and Tolsdorf, Jan and Wiefling, Stephan and Iacono, Luigi Lo and Schmitt, Hartmut},
  booktitle={Human Factors in Privacy Research},
  pages={83--113},
  year={2023},
  publisher={Springer International Publishing Cham},
  doi={10.1007/978-3-031-28643-8_5}
}

@article{glaser1965constant,
  title={The constant comparative method of qualitative analysis},
  author={Glaser, Barney G},
  journal={Social problems},
  volume={12},
  number={4},
  pages={436--445},
  year={1965},
  publisher={Oxford University Press Oxford, UK},
  doi={10.2307/798843}
}

@article{hevner2007three,
  title={A three cycle view of design science research},
  author={Hevner, Alan R},
  journal={Scandinavian journal of information systems},
  volume={19},
  number={2},
  pages={4},
  year={2007}
}

@article{zeng1991logic,
  title={On the logic of design},
  author={Zeng, Y and Cheng, GD},
  journal={Design Studies},
  volume={12},
  number={3},
  pages={137--141},
  year={1991},
  publisher={Elsevier},
  url={https://doi.org/10.1016/0142-694X(91)90022-O}
}

@article{tractinsky2000beautiful,
  title={What is beautiful is usable},
  author={Tractinsky, Noam and Katz, Adi S and Ikar, Dror},
  journal={Interacting with computers},
  volume={13},
  number={2},
  pages={127--145},
  year={2000},
  publisher={Oxford University Press Oxford, UK},
  url={https://doi.org/10.1016/S0953-5438(00)00031-X}
}

@article{obar2020biggest,
  title={The biggest lie on the internet: Ignoring the privacy policies and terms of service policies of social networking services},
  author={Obar, Jonathan A and Oeldorf-Hirsch, Anne},
  journal={Information, Communication \& Society},
  volume={23},
  number={1},
  pages={128--147},
  year={2020},
  publisher={Taylor \& Francis},
  url={https://doi.org/10.1080/1369118X.2018.1486870}
}

@inproceedings{mcdonald2009comparative,
  title={A comparative study of online privacy policies and formats},
  author={McDonald, Aleecia M and Reeder, Robert W and Kelley, Patrick Gage and Cranor, Lorrie Faith},
  booktitle={International Symposium on Privacy Enhancing Technologies Symposium},
  pages={37--55},
  year={2009},
  organization={Springer},
  url={https://doi.org/10.1007/978-3-642-03168-7_3}
}

@inproceedings{bhatia2016theory,
  title={A theory of vagueness and privacy risk perception},
  author={Bhatia, Jaspreet and Breaux, Travis D and Reidenberg, Joel R and Norton, Thomas B},
  booktitle={2016 IEEE 24th International Requirements Engineering Conference (RE)},
  pages={26--35},
  year={2016},
  organization={IEEE},
  url={https://doi.org/10.1109/RE.2016.20}
}

@article{rubinstein2013privacy,
  title={Privacy by design: A counterfactual analysis of Google and Facebook privacy incidents},
  author={Rubinstein, Ira S and Good, Nathaniel},
  journal={Berkeley Tech. LJ},
  volume={28},
  pages={1333},
  year={2013},
  publisher={HeinOnline},

}

@inproceedings{gray2024legal,
  title={Legal Trouble?: UX Practitioners' Engagement with Law and Regulation},
  author={Gray, Colin M and Gairola, Ritika and Boucaud, Nayah and Hashmi, Maliha and Chivukula, Shruthi Sai and Menon, Ambika R and Duane, Ja-Nae},
  booktitle={Companion Publication of the 2024 ACM Designing Interactive Systems Conference},
  pages={106--110},
  year={2024},
  url = {https://doi.org/10.1145/3656156.3663698}
}

@inproceedings{tran2025dark,
  title={Dark Patterns in the Opt-Out Process and Compliance with the California Consumer Privacy Act (CCPA)},
  author={Tran, Van Hong and Mehrotra, Aarushi and Sharma, Ranya and Chetty, Marshini and Feamster, Nick and Frankenreiter, Jens and Strahilevitz, Lior},
  booktitle={Proceedings of the 2025 CHI Conference on Human Factors in Computing Systems},
  pages={1--25},
  year={2025},
  url={https://doi.org/10.1145/3706598.3714138}
}

@inproceedings{shi202550,
  title={50 shades of deceptive patterns: A unified taxonomy, multimodal detection, and security implications},
  author={Shi, Zewei and Sun, Ruoxi and Chen, Jieshan and Sun, Jiamou and Xue, Minhui and Gao, Yansong and Liu, Feng and Yuan, Xingliang},
  booktitle={Proceedings of the ACM on Web Conference 2025},
  pages={978--989},
  year={2025},
  url={https://doi.org/10.1145/3696410.3714593}
}

@inproceedings{patil2014,
title = {Reflection or action? how feedback and control affect location sharing decisions},
author = {Patil, Sameer and Schlegel, Roman and Kapadia, Apu and Lee, Adam J.},
isbn = {9781450324731},
publisher = {Association for Computing Machinery},
address = {New York, NY, USA},
booktitle = {Proceedings of the SIGCHI Conference on Human Factors in Computing Systems},
pages = {101–110},
numpages = {10},
location = {Toronto, Ontario, Canada},
series = {CHI '14},
year = {2014},
url = {https://doi.org/10.1145/2556288.2557121},
}

@article{jiang2021location,
  title={Location privacy-preserving mechanisms in location-based services: A comprehensive survey},
  author={Jiang, Hongbo and Li, Jie and Zhao, Ping and Zeng, Fanzi and Xiao, Zhu and Iyengar, Arun},
  journal={ACM Computing Surveys (CSUR)},
  volume={54},
  number={1},
  pages={1--36},
  year={2021},
  publisher={ACM New York, NY, USA},
  url={https://doi.org/10.1145/3423165}
}

@inproceedings{liu2014reconciling,
  title={Reconciling mobile app privacy and usability on smartphones: Could user privacy profiles help?},
  author={Liu, Bin and Lin, Jialiu and Sadeh, Norman},
  booktitle={Proceedings of the 23rd international conference on World wide web},
  pages={201--212},
  year={2014},
  url={https://doi.org/10.1145/2566486.2568035}
}

@article{liu2016privacy,
  title={When privacy meets usability: Unobtrusive privacy permission recommendation system for mobile apps based on crowdsourcing},
  author={Liu, Rui and Cao, Jiannong and Zhang, Kehuan and Gao, Wenyu and Liang, Junbin and Yang, Lei},
  journal={IEEE Transactions on Services Computing},
  volume={11},
  number={5},
  pages={864--878},
  year={2016},
  publisher={IEEE},
  url={https://doi.org/10.1109/TSC.2016.2605089}
}

@inproceedings{obermeyer2019bias,
  title={Dissecting racial bias in an algorithm that guides health decisions for 70 million people},
  author={Obermeyer, Ziad and Mullainathan, Sendhil},
  booktitle={Proceedings of the Conference on Fairness, Accountability, and Transparency},
  pages={89--89},
  year={2019},
  url={https://doi.org/10.1145/3287560.3287593}
}

@inproceedings{chen2019bias,
  title={Correcting for recency bias in job recommendation},
  author={Chen, R. C. and Ai, Q. and Jayasinghe, G. and Croft, W. B.},
  booktitle={Proceedings of the 28th ACM International Conference on Information and Knowledge Management},
  pages={2185--2188},
  year={2019},
  month={November},
  organization={ACM},
  url={https://doi.org/10.1145/3357384.3358131}
}

@inproceedings{hajian2016bias,
  title={Algorithmic bias: From discrimination discovery to fairness-aware data mining},
  author={Hajian, Sara and Bonchi, Francesco and Castillo, Carlos},
  booktitle={Proceedings of the 22nd ACM SIGKDD International Conference on Knowledge Discovery and Data Mining},
  pages={2125--2126},
  year={2016},
  month={August},
  organization={ACM},
  url={https://doi.org/10.1145/2939672.2945386}
}

@article{strakova2021cookie,
  title={Human-computer interaction in the context of GDPR: How web users perceive and respond to blocking vs. non-blocking pop-ups},
  author={Straková, Sandra},
  journal={Master’s Thesis, Tilburg University},
  year={2021}
}

@misc{nyquist2023cookie,
  title={Cookies, GDPR and Dark Patterns: Effect on Consumer Privacy},
  author={Nyquist, Fredrik and Hildebrand, Tobias},
  howpublished={Bachelor's Thesis, Blekinge Institute of Technology},
  year={2023}
}

@inproceedings{felt2012dis,
  title={How to ask for permission},
  author={Felt, Adrienne Porter and Egelman, Serge and Finifter, Matthew and Akhawe, Devdatta and Wagner, David},
  booktitle={Proceedings of the 7th USENIX Conference on Hot Topics in Security (HotSec)},
  year={2012},
  organization={USENIX Association},
  url={https://www.usenix.org/conference/hotsec12/workshop-program/presentation/felt}
}

@article{adjerid2016impact,
  title={The impact of privacy regulation and technology incentives: The case of health information exchanges},
  author={Adjerid, Idris and Acquisti, Alessandro and Telang, Rahul and Padman, Rema and Adler-Milstein, Julia},
  journal={Management Science},
  volume={62},
  number={4},
  pages={1042--1063},
  year={2016},
  publisher={INFORMS},
  url={https://doi.org/10.1287/mnsc.2015.2194}
}

@article{acquisti2017nudges,
  title={Nudges for privacy and security: Understanding and assisting users’ choices online},
  author={Acquisti, Alessandro and Adjerid, Idris and Balebako, Rebecca and Brandimarte, Laura and Cranor, Lorrie Faith and Komanduri, Saranga and Leon, Pedro Giovanni and Sadeh, Norman and Schaub, Florian and Sleeper, Manya and others},
  journal={ACM Computing Surveys (CSUR)},
  volume={50},
  number={3},
  pages={1--41},
  year={2017},
  publisher={ACM New York, NY, USA},
  url={https://doi.org/10.1145/3054926}
}

@article{hong2001infrastructure,
  title={An infrastructure approach to context-aware computing},
  author={Hong, Jason I and Landay, James A},
  journal={Human--computer interaction},
  volume={16},
  number={2-4},
  pages={287--303},
  year={2001},
  publisher={Taylor \& Francis},
  url={https://doi.org/10.1207/S15327051HCI16234_11}
}

@inproceedings{wang2013privacy,
  title={Privacy nudges for social media: an exploratory Facebook study},
  author={Wang, Yang and Leon, Pedro Giovanni and Scott, Kevin and Chen, Xiaoxuan and Acquisti, Alessandro and Cranor, Lorrie Faith},
  booktitle={Proceedings of the 22nd international conference on world wide web},
  pages={763--770},
  year={2013},
  url={https://doi.org/10.1145/2487788.2488038}
}

@article{jones2019determinants,
  title={Determinants for successful agile collaboration between UX designers and software developers in a complex organisation},
  author={Jones, Alexander and Thoma, Volker},
  journal={International Journal of Human--Computer Interaction},
  volume={35},
  number={20},
  pages={1914--1935},
  year={2019},
  publisher={Taylor \& Francis},
  url={https://doi.org/10.1080/10447318.2019.1587856}
}

@article{koops2014privacy,
  title={Privacy regulation cannot be hardcoded. A critical comment on the ‘privacy by design’provision in data-protection law},
  author={Koops, Bert-Jaap and Leenes, Ronald},
  journal={International Review of Law, Computers \& Technology},
  volume={28},
  number={2},
  pages={159--171},
  year={2014},
  publisher={Taylor \& Francis},
  url={https://doi.org/10.1080/13600869.2013.801589}
}

@misc{chen2017,
  author       = {Chen, T. W.},
  title        = {Can Overt Personalization and Transparency Enhance User Experience of Personalized Mobile Services?},
  year         = {2017},
  howpublished = {\url{https://etda.libraries.psu.edu/catalog/13848tzc144}},
  note         = {ETDA Thesis, Pennsylvania State University}
}

@inproceedings{pattan2009study,
  title={Study of usability of security and privacy in context aware mobile applications},
  author={Pattan, Neha and Madamanchi, Deepthi},
  booktitle={International Conference on Mobile Computing, Applications, and Services},
  pages={326--330},
  year={2009},
  organization={Springer},
  url={https://doi.org/10.1007/978-3-642-12607-9_21}
}

@inproceedings{schafer2023,
author = {Sch\"{a}fer, Ren\'{e} and Preuschoff, Paul Miles and Borchers, Jan},
title = {Investigating Visual Countermeasures Against Dark Patterns in User Interfaces},

publisher = {Association for Computing Machinery},
address = {New York, NY, USA},
booktitle = {Proceedings of Mensch Und Computer 2023},
pages = {161–172},
numpages = {12},
location = {Rapperswil, Switzerland},
series = {MuC '23},
year = {2023},
url = {https://doi.org/10.1145/3603555.3603563}
}

@article{aung2024stupid,
  title={" What a stupid way to do business": Towards an Understanding of Older Adults' Perceptions of Deceptive Patterns and Ways to Develop Resistance},
  author={Aung, Kalya Win and Soubutts, Ewan and Singh, Aneesha},
  journal={Proceedings of the ACM on Human-Computer Interaction},
  volume={8},
  number={CHI PLAY},
  pages={1--31},
  year={2024},
  publisher={ACM New York, NY, USA},
  url={https://doi.org/10.1145/3677113}
}

@inproceedings{lukoff2021can,
  title={What can CHI do about dark patterns?},
  author={Lukoff, Kai and Hiniker, Alexis and Gray, Colin M and Mathur, Arunesh and Chivukula, Shruthi Sai},
  booktitle={Extended abstracts of the 2021 chi conference on human factors in computing systems},
  pages={1--6},
  year={2021},
  url={https://doi.org/10.1145/3411763.3441360}
}

@article{lu2024awareness,
  title={From awareness to action: Exploring end-user empowerment interventions for dark patterns in ux},
  author={Lu, Yuwen and Zhang, Chao and Yang, Yuewen and Yao, Yaxing and Li, Toby Jia-Jun},
  journal={Proceedings of the ACM on Human-Computer Interaction},
  volume={8},
  number={CSCW1},
  pages={1--41},
  year={2024},
  publisher={ACM New York, NY, USA},
  url={https://doi.org/10.1145/3637336}
}

@inproceedings{lara2016enhancing,
  title={Enhancing privacy notice applications through interaction design},
  author={Lara, Enrique S{\'a}nchez and Murillo, Sandra R and S{\'a}nchez, J Alfredo},
  booktitle={2016 4th International Conference in Software Engineering Research and Innovation (CONISOFT)},
  pages={1--8},
  year={2016},
  organization={IEEE},
  url={https://doi.org/10.1109/CONISOFT.2016.9}
}

@inproceedings{ko2007usability,
  title={Usability Enhanced Privacy Protection System Based on Users' Responses},
  author={Ko, Han-Gyu and Kim, Seung-Hyun and Jin, Seung-Hun},
  booktitle={2007 IEEE International Symposium on Consumer Electronics},
  pages={1--6},
  year={2007},
  organization={IEEE},
  url={https://doi.org/10.1109/ISCE.2007.4382188}
}

@inproceedings{earp2016had,
  title={" I had no idea this was a thing" on the importance of understanding the user experience of personalized transparency tools},
  author={Earp, Julia and Staddon, Jessica},
  booktitle={Proceedings of the 6th Workshop on Socio-Technical Aspects in Security and Trust},
  pages={79--86},
  year={2016},
  url={https://doi.org/10.1145/3046055.3046062}
}

@inproceedings{micallef2017stop,
  title={Stop annoying me! an empirical investigation of the usability of app privacy notifications},
  author={Micallef, Nicholas and Just, Mike and Baillie, Lynne and Alharby, Maher},
  booktitle={Proceedings of the 29th Australian Conference on Computer-Human Interaction},
  pages={371--375},
  year={2017},
  url={https://doi.org/10.1145/3152771.3156139}
}

@book{iachello2007end,
  title={End-user privacy in human-computer interaction},
  author={Iachello, Giovanni},
  year={2007},
  publisher={Now Publishers Inc}
}

@article{zloteanu2018digital,
  title={Digital identity: The effect of trust and reputation information on user judgement in the sharing economy},
  author={Zloteanu, Mircea and Harvey, Nigel and Tuckett, David and Livan, Giacomo},
  journal={PloS one},
  volume={13},
  number={12},
  pages={e0209071},
  year={2018},
  publisher={Public Library of Science San Francisco, CA USA},
  url={https://doi.org/10.1371/journal.pone.0209071}
}

@inproceedings{assal2015s,
  title={What's the deal with privacy apps? A comprehensive exploration of user perception and usability},
  author={Assal, Hala and Hurtado, Stephanie and Imran, Ahsan and Chiasson, Sonia},
  booktitle={Proceedings of the 14th International Conference on Mobile and Ubiquitous Multimedia},
  pages={25--36},
  year={2015},
  url={https://doi.org/10.1145/2836041.2836044}
}

@article{martini2022making,
  title={Making choice meaningful--tackling dark patterns in cookie and consent banners through european data privacy law},
  author={Martini, Prof and Drews, Christian and others},
  journal={Available at SSRN 4257979},
  year={2022},
  url={https://dx.doi.org/10.2139/ssrn.4257979}
}

@inproceedings{schafer2024fighting,
  title={Fighting malicious designs: towards visual countermeasures against dark patterns},
  author={Sch{\"a}fer, Ren{\'e} and Preuschoff, Paul Miles and R{\"o}pke, Ren{\'e} and Sahabi, Sarah and Borchers, Jan},
  booktitle={Proceedings of the 2024 CHI Conference on Human Factors in Computing Systems},
  pages={1--13},
  year={2024},
  url={https://doi.org/10.1145/3613904.3642661}
}

@article{jovanovic2022vortex,
  title={VoRtex Metaverse platform for gamified collaborative learning},
  author={Jovanovi{\'c}, Aleksandar and Milosavljevi{\'c}, Aleksandar},
  journal={Electronics},
  volume={11},
  number={3},
  pages={317},
  year={2022},
  publisher={MDPI},
  url={https://doi.org/10.3390/electronics11030317}
}

@inproceedings{drake2018designing,
  title={Designing a User-Experience-First, Privacy-Respectful, high-security mutual-multifactor authentication solution},
  author={Drake, Chris and Gauravaram, Praveen},
  booktitle={International Symposium on Security in Computing and Communication},
  pages={183--210},
  year={2018},
  organization={Springer},
  url={https://doi.org/10.1007/978-981-13-5826-5_14}
}

@inproceedings{hirnschall2018learning,
  title={Learning user preferences to incentivize exploration in the sharing economy},
  author={Hirnschall, Christoph and Singla, Adish and Tschiatschek, Sebastian and Krause, Andreas},
  booktitle={Proceedings of the AAAI Conference on Artificial Intelligence},
  volume={32},
  number={1},
  year={2018},
  url={https://doi.org/10.1609/aaai.v32i1.11874}
}

@inproceedings{moore2024negative,
  title={Negative effects of social triggers on user security and privacy behaviors},
  author={Moore, Lachlan and Mori, Tatsuya and Hasegawa, Ayako A},
  booktitle={Twentieth Symposium on Usable Privacy and Security (SOUPS 2024)},
  pages={605--622},
  year={2024},
url = {https://www.usenix.org/conference/soups2024/presentation/moore}
}

@misc{ACM2020OnlinePersuasion,
  author       = {{Netherlands Authority for Consumers and Markets}},
  title        = {Guidelines on the Protection of the Online Consumer: Boundaries of Online Persuasion},
  year         = {2020},
  howpublished = {Policy guideline},
}
\balance

\onecolumn
\appendix

\section*{Search Queries}
\begin{table*}[h]
    \centering
    \scriptsize
    \renewcommand{\arraystretch}{1.2}
    \caption{Search queries used in Google Scholar for the SLR.}
    \resizebox{0.99\linewidth}{!}{
    \begin{tabular}{l|p{14cm}} 
    \hline
        \textbf{Query} & \textbf{Search String (title only)} \\
        \hline
         SS1 & (``UI/UX design'' OR ``User interface design'' OR ``User experience design'' OR ``Interaction design'' OR ``Bright patterns'' OR ``Dark patterns'' OR ``Deceptive patterns'') 
        AND (``Priva*'' OR ``Data privacy'' OR ``Privacy-preserving'' OR ``Privacy-enhancing'' OR ``Data protection'' OR ``GDPR'') \\
        SS2 & (``UI/UX design'' OR ``User interface design'' OR ``User experience design'' OR ``Interaction design'' OR ``Bright patterns'' OR ``Dark patterns'' OR ``Deceptive patterns'') 
        AND (``CCPA'' OR ``Legal compliance'' OR ``Privacy Considerations'' OR ``Privacy by design'' OR ``User consent'' OR ``Transparency'') \\
        SS3 & (``UI/UX design'' OR ``User interface design'' OR ``User experience design'' OR ``Interaction design'' OR ``Bright patterns'' OR ``Dark patterns'' OR ``Deceptive patterns'') 
        AND (``User control'' OR ``User rights'' OR ``Data sharing'' OR ``Data minimization'' OR ``Data security'' OR ``User trust'' OR ``ePrivacy Directive'') \\
        SS4 & (``Privacy Design patterns'' OR ``Human-computer interaction'' OR ``HCI'' OR ``Usability'') 
        AND (``Priva*'' OR ``Data privacy'' OR ``Privacy-preserving'' OR ``Privacy-enhancing'' OR ``Data protection'' OR ``GDPR'') \\
         SS5 & (``Privacy Design patterns'' OR ``Human-computer interaction'' OR ``HCI'' OR ``Usability'') 
        AND (``CCPA'' OR ``Legal compliance'' OR ``Privacy Considerations'' OR ``Privacy by design'' OR ``User consent'' OR ``Transparency'') \\
        SS6 & (``Privacy Design patterns'' OR ``Human-computer interaction'' OR ``HCI'' OR ``Usability'') 
        AND (``User control'' OR ``User rights'' OR ``Data sharing'' OR ``Data minimization'' OR ``Data security'' OR ``User trust'' OR ``ePrivacy Directive'') \\
         SS7 & (``User experience'' OR ``UI elements'' OR ``Ethical design'' OR ``User-centered design'' OR ``UX research'') 
        AND (``Priva*'' OR ``Data privacy'' OR ``Privacy-preserving'' OR ``Privacy-enhancing'' OR ``Data protection'' OR ``GDPR'') \\
        SS8 & (``User experience'' OR ``UI elements'' OR ``Ethical design'' OR ``User-centered design'' OR ``UX research'') 
        AND (``CCPA'' OR ``Legal compliance'' OR ``Privacy Considerations'' OR ``Privacy by design'' OR ``User consent'' OR ``Transparency'') \\
        SS9 & (``User experience'' OR ``UI elements'' OR ``Ethical design'' OR ``User-centered design'' OR ``UX research'') 
        AND (``User control'' OR ``User rights'' OR ``Data sharing'' OR ``Data minimization'' OR ``Data security'' OR ``User trust'' OR ``ePrivacy Directive'') \\
    \end{tabular}
    }
    \label{tab:search_queries}
\end{table*}

\section*{Interview Guide}
\label{sec:interview_guide}
\small
\textbf{Disclaimer.} By taking part in this interview, you agree for the audio of the interview to be recorded and transcribed for further analysis. The interview transcripts will not be shared with anyone outside of our immediate research team. You also agree for direct quotes from the interview to be used for publication, and that such quotes will be attributed in a pseudonymized form. No PII or any other personal data will be shared or attributed. Please confirm your consent to these terms. \newline

\textbf{Familiarity with Privacy in General} 
\begin{enumerate}
    \item[1.] *Could you briefly describe your experience with UI/UX design (development)? What types of products or projects have you primarily worked on?
    
    
\end{enumerate}

\textbf{I. Identifying Privacy Considerations} 
\begin{enumerate}
        \item[2.] How familiar are you with privacy concepts in general? How do you think these concepts might relate to your work in UI/UX design? 
    
        \item[3.] What role, if any, does privacy play in your design work? Could you describe some contexts where privacy may come up? 
        \item[4.] What factors influence your perspective on privacy in design? (e.g. opinions, company policies, or industry standards) How do you typically approach it when it arises?
\end{enumerate}

\textbf{II. Key Factors in Addressing Privacy in Design}
\begin{enumerate}
        \item[5.] What factors are most crucial to you when addressing privacy in design? 
        \begin{itemize}
            \item Why are these aspects important in your decision-making?
            \item Could you share an example of how these factors play a role in shaping your design choices?
        \end{itemize} 
        \item[6.] What primary challenges or barriers have you faced in addressing privacy considerations? If possible, could you describe a specific scenario?
\end{enumerate}

\textbf{III. Strategies and Approaches for Better Integration}
\begin{enumerate}
        \item[7.] When faced with design requests that might raise privacy concerns, what strategies do you use to address these challenges in your own work? 
        \begin{itemize}
            \item Could you describe any strategies you think would be effective, even if you haven’t encountered such scenarios yet?
        \end{itemize}
        \item[8.] Could you describe an approach you might take to ensure privacy and other ethical considerations are respected?
        \item[9.] What experience do you have, if any, with using privacy-preserving patterns in your design? 
        \begin{itemize}
            \item If not, what resources or support would help you on that? (e.g. tools, frameworks, or design references)
        \end{itemize}
        \item[10.] What would make it easier for you to incorporate privacy-preserving practices into your design process?
        \item [11.] Placeholder for another question, which will be asking for Artifact Feedback. [Based on the artifacts being collected, identified in the current literature, you will be asked for feedback on their effectiveness in supporting privacy-preserving design and address gaps that are currently present.]
\end{enumerate}

\textbf{Looking Forward}
\begin{enumerate}
    \item[12.] *In your opinion, how do you think privacy considerations in UI/UX will evolve over the next few years? What do you think will drive these changes?
\end{enumerate}

\textbf{Other}
\begin{enumerate}
    \item[13.] Is there anything we haven’t covered that you think is important to mention on this topic?
    \item[14.] Can you suggest any colleagues or other professionals who may have valuable insights on privacy-preserving design patterns?
\end{enumerate}

\vspace{10pt}
\noindent\rule{18.1cm}{0.4pt}
\vspace{10pt}

\section*{Survey Questions}
\label{sec:survey_questions}
\small

\textbf{Disclaimer.}
By taking part in this survey, you agree for your responses to be used for further analysis. They will not be shared with anyone outside of our immediate research team. No PII or any other personal data will be shared or attributed. Please confirm your consent to these terms. \newline

\textbf{Personal Information}

\begin{enumerate}
    \item[1.] What is your email address? [\texttt{Open Text Response}]
    \item[2.] What is your first name? [\texttt{Open Text Response}]
    \item[3.] What is your last name? [\texttt{Open Text Response}]
    \item[4.] What is your gender? [\texttt{Female | Male | Non-binary | Prefer not to say}]
    \item[5.] What is the highest level of formal education that you have completed or are currently pursuing? [\texttt{High School | Bachelor's or equivalent | Master's or equivalent | Doctorate or equivalent}]
\end{enumerate}

\textbf{Company Information}

\begin{enumerate}
    \item[6.] What is the industry domain of the company you are employed at? [\texttt{Automotive | Manufacturing | Retail | Finance | Information Technology | Healthcare | Energy | Electronics | Other: [Open Text Response]}]
    \item[7.] What is the size of the company you are employed at? [\texttt{Micro (1-9 employees) | Small (10-49 employees) | Medium (50-249 employees) | Large (250 employees or more) | N/A}]
    \item[8.] What country do you work in? [\texttt{Dropdown list of countries}]
\end{enumerate}

\textbf{Professional Information}

\begin{enumerate}
    \item[9.] What is your official position? [\texttt{Open Text Response}]
    \item[10.] How many years of professional experience do you have? [\texttt{1 - 3 | 3 - 5 | 5 - 10 | 10 - 20 | 20+}]
    \item[11.] What best describes your experience with design-related tasks or projects? [\texttt{Limited experience (e.g., occasional involvement in design-related tasks or collaborations) | Moderate experience (e.g., contributing to some design elements in projects) | Extensive experience (e.g., designing as a core part of your role) | Expert-level experience (e.g., significant expertise in design or leading major design projects)}]
\end{enumerate}

\  \\
\underline{\textbf{NOTE}}: [Unless otherwise noted, for questions 12-30, the answer options are on the Likert scale of [\texttt{strongly disagree | disagree | neutral | agree | strongly agree}]. 

\textbf{Catalog Feedback}

\begin{enumerate}
    \item[12.] The content is interesting and engaging. 
    \item[13.] The content is presented in a clear and understandable way.
    \item[14.] The research goals are well aligned with the content.
    \item[15.] Key points and significant ideas are emphasized with an appropriate level of detail.
    \item[16.] The content follows a logical structure.
    \item[17.] If not, what would make the structure more logical? [\texttt{Open Text Response}]
\end{enumerate}

\textbf{Perceived Outcomes}

\begin{enumerate}
    \item[18.] Claims are supported by evidence or logical argument.
    \item[19.] The design of the learning content is visually appealing and does not distract from the content.
    \item[20.] I can name and define various Privacy-preserving patterns.
    \item[21.] I understand the main principles of how the covered categories work.
    \item[22.] I would be able to apply the recommendations in my design easily.
    \item[23.] If not, what would contribute to easier integration of the recommendations in your design work? [\texttt{Open Text Response}]
    \item[24.] I can more confidently communicate on the topics of privacy considerations in UI/UX design.
    \item[25.] I would be able to evaluate whether a pattern should be privacy-preserving in a specific user scenario.
    \item[26.] Overall, I now have a better understanding of Privacy-preserving UI/UX design than before.
    \item[27.] The acquired knowledge helped me to better understand the technical side of data privacy.
    \item[28.] The acquired knowledge will help me more efficiently communicate with designers/developers on the topics of data privacy.
\end{enumerate}

\textbf{Final Thoughts}

\begin{enumerate}
    \item[29.] The acquired knowledge will help me in my work when facing challenges with privacy topics.
    \item[30.] I would recommend this learning content to my colleagues.
    \item[31.] What did you like about the pattern catalog? [\texttt{Open Text Response}]
    \item[32.] How could the pattern catalog be improved? [\texttt{Open Text Response}]
\end{enumerate}

\newpage
\section*{Literature Sources}

\begin{table}[ht!]
\centering
\caption{A complete listing of the included sources in the SLR.}
\resizebox*{!}{0.91\textheight}{
\begin{tabular}{c|p{1.15\linewidth}|c}
\textbf{Year} & \textbf{Title} & \textbf{Source} \\ \hline
- & PrivacyPatterns.org: A Collection of Privacy Design Patterns & \cite{privacypatterns} \\
- & A Collection of Ethical and Respectful Design Patterns & \cite{brightpatterns} \\
- & Design Patterns for Ethical Tech & \cite{decentpatterns} \\
- & Catalogue of Digital Rights & \cite{projectsbyif2025} \\
2025 & Dark Patterns in the Opt-Out Process and Compliance with the California Consumer Privacy Act (CCPA) & \cite{tran2025dark} \\
2025 & 50 shades of deceptive patterns: A unified taxonomy, multimodal detection, and security implications & \cite{shi202550} \\
2024 & Navigating the Gray: Design Practitioners' Perceptions Toward the Implementation of Privacy Dark Patterns & \cite{zhangkennedy2024} \\
2024 & Challenges of Usable Privacy & \cite{fischer2024challenges} \\
2024 & Data Sharing and Website Competition: The Role of Dark Patterns & \cite{fradkin2024ssrn} \\
2024 & Ethical Design for Data Privacy and User Privacy Awareness in the Metaverse & \cite{prillard2024onboarding} \\
2024 & Learning from the Dark Side About How (Not) to Engineer Privacy: Analysis of Dark Patterns Taxonomies from an ISO 29100 Perspective & \cite{valoggia2024} \\
2024 & Legal Trouble?: UX Practitioners' Engagement with Law and Regulation & \cite{gray2024legal} \\
2024 & Beyond Dark Patterns: A Concept-Based Framework for Ethical Software Design & \cite{caragay2024} \\
2024 & Embedding Privacy First Human Centric in User Experience for Mobile Applications & \cite{phokela2024location} \\
2024 & ``What a stupid way to do busines'': Towards an Understanding of Older Adults' Perceptions of Deceptive Patterns and Ways to Develop Resistance & \cite{aung2024stupid} \\
2024 & From awareness to action: Exploring end-user empowerment interventions for dark patterns in ux & \cite{lu2024awareness} \\
2024 & Fighting malicious designs: towards visual countermeasures against dark patterns & \cite{schafer2024fighting} \\
2024 & Negative effects of social triggers on user security and privacy behaviors & \cite{moore2024negative} \\
2023 & How Americans view data privacy & \cite{mcclain2023americans} \\
2023 & Defining and identifying attention capture deceptive designs in digital interfaces & \cite{Roffarello2023} \\
2023 & Achieving usable security and privacy through Human-Centered Design & \cite{groen2023achieving} \\
2023 & Ethical tensions in UX design practice: exploring the fine line between persuasion and manipulation in online interfaces & \cite{sanchez2023ethical} \\
2023 & `Don’t Annoy Me With Privacy Decisions!'—Designing Privacy-Preserving User Interfaces for SSI Wallets on Smartphones & \cite{teuschel2023don} \\
2023 & Cookies, GDPR and Dark Patterns: Effect on Consumer Privacy & \cite{nyquist2023cookie} \\
2023 & Investigating Visual Countermeasures Against Dark Patterns in User Interfaces & \cite{schafer2023} \\
2022 & Rationalizing dark patterns: Examining the process of designing privacy UX through speculative enactments & \cite{nelissen2022rationalizing} \\
2022 & Evaluating the Usability of Privacy Choice Mechanisms & \cite{habib2022usability} \\
2022 & Redress for dark patterns privacy harms? A case study on consent interactions & \cite{gunawan2022redress} \\
2022 & Exploring the experience of ethical tensions and the role of community in UX practice & \cite{pillai2022} \\
2022 & Dark patterns: The case for regulatory pluralism between the European Unions consumer and data protection regimes & \cite{leiser2022dark} \\
2022 & Implications of XR on Privacy, Security and Behaviour: Insights from Experts & \cite{abraham2022implications} \\
2022 & Making choice meaningful--tackling dark patterns in cookie and consent banners through european data privacy law & \cite{martini2022making} \\
2022 & VoRtex Metaverse platform for gamified collaborative learning & \cite{jovanovic2022vortex} \\
2021 & ``I am Definitely Manipulated, Even When I am Aware of it. It's Ridiculous!'' -- Dark Patterns from the End-User Perspective & \cite{bongard2021definitely} \\
2021 & What Makes a Dark Pattern... Dark? Design Attributes, Normative Considerations, and Measurement Methods & \cite{mathur2021} \\
2021 & Building a privacy oriented UI and UX design: an introduction to its foundations and potential developments & \cite{parrilli2021building} \\
2021 & Shining a light on dark patterns & \cite{luguri2021shining} \\
2021 & User tracking in the post-cookie era: How websites bypass GDPR consent to track users & \cite{papadogiannakis2021cookie} \\
2021 & Privacy policies between perception and learning through legal design: Ideas for an educational chatbot combining rights' awareness, optimized user experience and training efficacy & \cite{guida2021} \\
2021 & Toggles, dollar signs, and triangles: How to (in) effectively convey privacy choices with icons and link texts & \cite{habib2021toggles} \\
2021 & Location privacy-preserving mechanisms in location-based services: A comprehensive survey & \cite{jiang2021location} \\
2021 & Human-computer interaction in the context of GDPR: How web users perceive and respond to blocking vs. non-blocking pop-ups & \cite{strakova2021cookie} \\
2021 & What can CHI do about dark patterns? & \cite{lukoff2021can} \\
2020 & Cognitive biases, dark patterns, and the ‘privacy paradox’ & \cite{waldman2020} \\
2020 & Are dark patterns anticompetitive? & \cite{day2020} \\
2020 & Qualified Statewide Ballot Measures & \cite{california2020} \\
2020 & Guidelines on the Protection of the Online Consumer: Boundaries of Online Persuasion & \cite{ACM2020OnlinePersuasion} \\
2020 & Dark patterns after the GDPR: Scraping consent pop-ups and demonstrating their influence & \cite{nouwens2020consentpopup} \\
2020 & The fertile dark matter of privacy takes on the dark patterns of surveillance & \cite{mulligan2020} \\
2020 & The biggest lie on the internet: Ignoring the privacy policies and terms of service policies of social networking services & \cite{obar2020biggest} \\
2020 & Dark patterns--An end user perspective & \cite{maier2020} \\
2019 & Bringing design to the privacy table: Broadening “design” in “privacy by design” through the lens of HCI & \cite{wong2019} \\
2019 & Dark patterns at scale: Findings from a crawl of 11K shopping websites & \cite{mathur2019} \\
2019 & Opt out of privacy or" go home": Understanding reluctant privacy behaviours through the FoMO-centric design paradigm & \cite{westin2019} \\
2019 & Shaping choices in the digital world. From dark patterns to data protection: the influence of ux/ui design on user empowerment & \cite{cnil2020} \\
2019 & Senators introduce bipartisan legislation to ban manipulative “Dark Patterns” & \cite{warner2019} \\
2019 & Ethical mediation in UX practice & \cite{gray2019ethical} \\
2019 & Dissecting racial bias in an algorithm that guides health decisions for 70 million people & \cite{obermeyer2019bias} \\
2019 & Correcting for recency bias in job recommendation & \cite{chen2019bias} \\
2019 & Determinants for successful agile collaboration between UX designers and software developers in a complex organisation & \cite{jones2019determinants} \\
2018 & The Dark (Patterns) Side of UX Design & \cite{gray2018uxside} \\
2018 & Dark Patterns & \cite{brignull2018} \\
2018 & Digital identity: The effect of trust and reputation information on user judgement in the sharing economy & \cite{zloteanu2018digital} \\
2018 & Designing a User-Experience-First, Privacy-Respectful, high-security mutual-multifactor authentication solution & \cite{drake2018designing} \\
2018 & Learning user preferences to incentivize exploration in the sharing economy & \cite{hirnschall2018learning} \\
2017 & Privacy Dashboards: Control and Understanding of Data through Usability and User Experience & \cite{vanGogh2017} \\
2017 & Privacy dark patterns in identity management & \cite{fritsch2017} \\
2017 & Nudges for privacy and security: Understanding and assisting users’ choices online & \cite{acquisti2017nudges} \\
2017 & Can Overt Personalization and Transparency Enhance User Experience of Personalized Mobile Services? & \cite{chen2017} \\
2017 & Stop annoying me! an empirical investigation of the usability of app privacy notifications & \cite{micallef2017stop} \\
2016 & Tales from the dark side: Privacy dark strategies and privacy dark patterns & \cite{bosch2016} \\
2016 & A usability study on the privacy policy visualization model & \cite{ghazinour2016ppvm} \\
2016 & A theory of vagueness and privacy risk perception & \cite{bhatia2016theory} \\
2016 & When privacy meets usability: Unobtrusive privacy permission recommendation system for mobile apps based on crowdsourcing & \cite{liu2016privacy} \\
2016 & Algorithmic bias: From discrimination discovery to fairness-aware data mining & \cite{hajian2016bias} \\
2016 & The impact of privacy regulation and technology incentives: The case of health information exchanges & \cite{adjerid2016impact} \\
2016 & Enhancing privacy notice applications through interaction design & \cite{lara2016enhancing} \\
2016 & "I had no idea this was a thing" on the importance of understanding the user experience of personalized transparency tools & \cite{earp2016had} \\
2015 & A design space for effective privacy notices & \cite{schaub2015design} \\
2015 & What's the deal with privacy apps? A comprehensive exploration of user perception and usability & \cite{assal2015s} \\
2014 & Privacy design strategies & \cite{hoepman2014} \\
2014 & Artifact evaluation in information systems design-science research–a holistic view & \cite{prat2014artifactevaluation} \\
2014 & Privacy dashboards: reconciling data-driven business models and privacy & \cite{Zimmermann2014dashboards} \\
2014 & Irresistible Apps: Motivational design patterns for apps, games, and web-based communities & \cite{lewis2014} \\
2014 & Reflection or action? how feedback and control affect location sharing decisions & \cite{patil2014} \\
2014 & Reconciling mobile app privacy and usability on smartphones: Could user privacy profiles help? & \cite{liu2014reconciling} \\
2014 & Privacy regulation cannot be hardcoded. A critical comment on the ‘privacy by design’ provision in data-protection law & \cite{koops2014privacy} \\
2013 & Privacy by design: A counterfactual analysis of Google and Facebook privacy incidents & \cite{rubinstein2013privacy} \\
2013 & Consent for all: revealing the hidden complexity of terms and conditions & \cite{luger2013consent} \\
2013 & Privacy nudges for social media: an exploratory Facebook study & \cite{wang2013privacy} \\
2012 & Privacy by Design and User Interfaces & \cite{cavoukian2012privacy} \\
2012 & The mismeasurement of privacy: using contextual integrity to reconsider privacy in HCI & \cite{barkhuus2012} \\
2012 & How to ask for permission & \cite{felt2012dis} \\
2009 & A comparative study of online privacy policies and formats & \cite{mcdonald2009comparative} \\
2009 & Study of usability of security and privacy in context aware mobile applications & \cite{pattan2009study} \\
2007 & Usability Enhanced Privacy Protection System Based on Users' Responses & \cite{ko2007usability} \\
2007 & End-user privacy in human-computer interaction & \cite{iachello2007end} \\
2004 & Privacy as contextual integrity & \cite{nissenbaum2004} \\
2004 & Strategies for reducing online privacy risks: Why consumers read (or don't read) online privacy notices & \cite{milne2004strategies} \\
2001 & An infrastructure approach to context-aware computing & \cite{hong2001infrastructure} \\
2000 & What is beautiful is usable & \cite{tractinsky2000beautiful} \\
1991 & On the logic of design & \cite{zeng1991logic} \\
\end{tabular}
}
\label{tab:slr}
\end{table}

\newpage
\section*{UI/UX Privacy Pattern Catalog}
\begin{figure*}[htbp]
    \centering
    \includegraphics[scale=0.2]{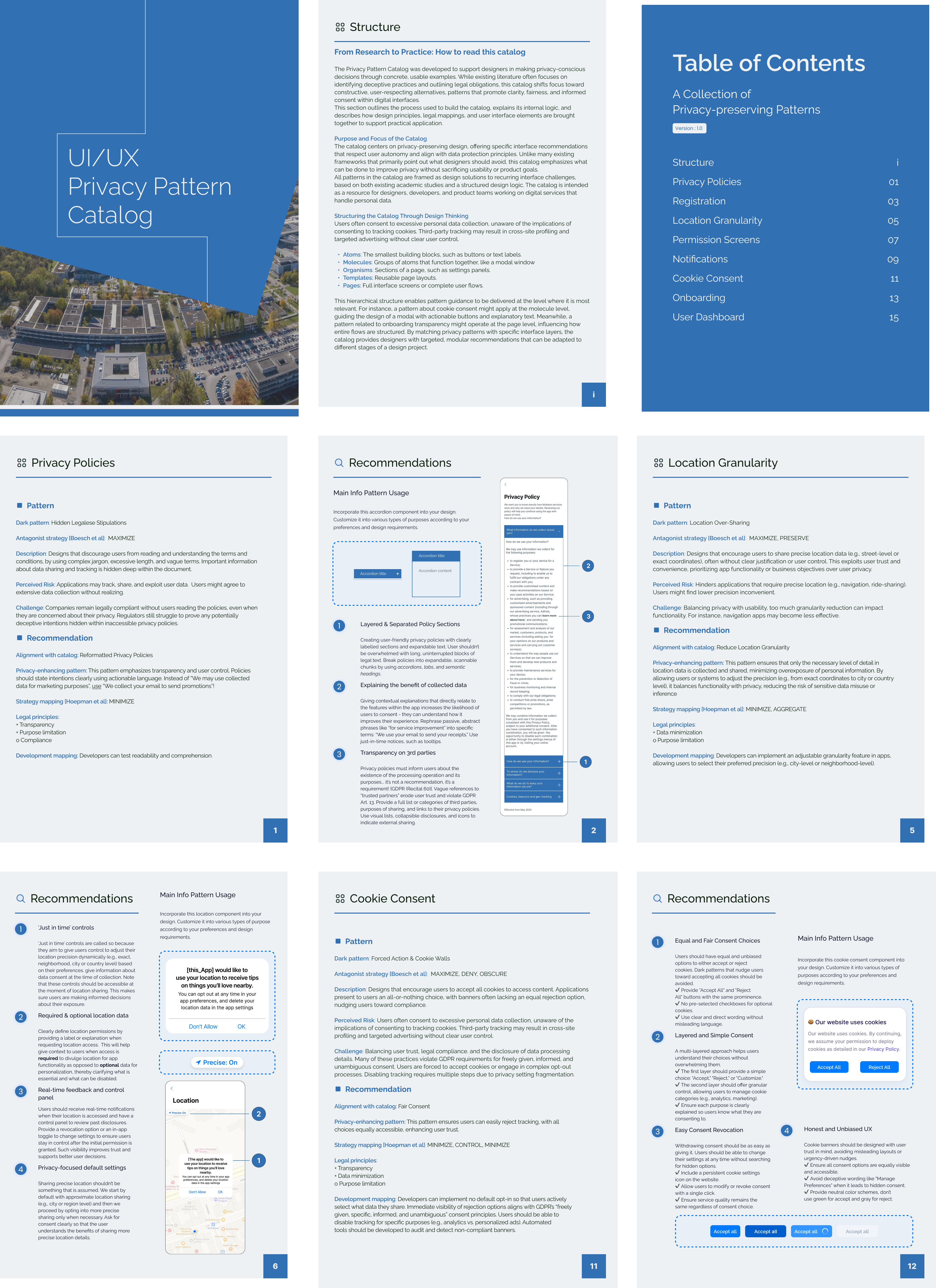}
    \caption{Selected pages from our \textit{UI/UX Privacy Pattern Catalog}.}
    \label{fig:catalog}
\end{figure*}



\newpage
\section*{Codebook}
\begin{figure}[ht!]
    \centering
    \includegraphics[width=0.99\linewidth]{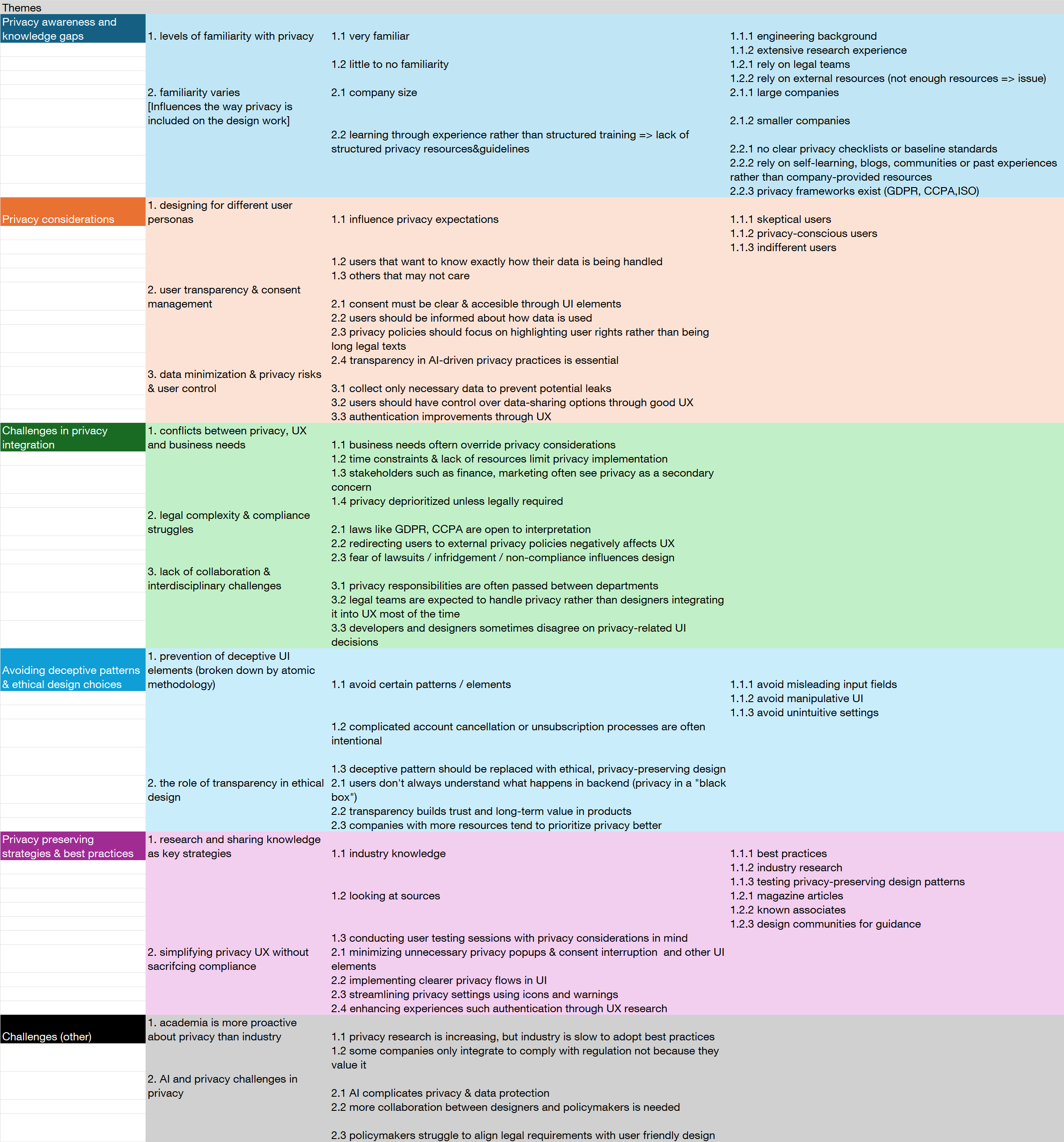}
    \caption{The complete codebook resulting from the thematic analysis of the interview study.}
    \label{fig:codebook}
\end{figure}

\end{document}